\documentclass[twocolumn,amsmath]{aastex701}

\usepackage{natbib}
\usepackage{graphicx}
\usepackage{dcolumn}
\usepackage{bm}
\usepackage{amsmath}
\usepackage{algorithm} 
\usepackage{algpseudocode} 
\usepackage{hyperref}
\usepackage{xspace}
\usepackage{multirow}

\def\simlt{\lower.5ex\hbox{$\; \buildrel < \over \sim \;$}}
\def\simgt{\lower.5ex\hbox{$\; \buildrel > \over \sim \;$}}
\ifx \change \undefined
  \newcommand{\change}[1]{{#1}}
\fi
\ifx \forcemathnospace \undefined
  \newcommand{\forcemathnospace}[1]{\ifmmode{\rm{#1}}\else {$\rm{#1}$}\fi\xspace}
\fi
\ifx \forcemath \undefined
  \newcommand{\forcemath}[1]{\ifmmode{\rm{#1}}\else {$\rm{#1}$}\fi\xspace}
\fi
\ifx \unitnospace \undefined
  \newcommand{\unitnospace}[1]{\forcemath{\rm #1}}
\fi
\ifx \unit \undefined
  \newcommand{\unit}[1]{\forcemath{\rm #1}}
\fi
\newcommand{\Aeff}{\forcemath{A_{\rm eff}}}
\newcommand{\Smin}{\forcemath{S_{\rm min}}}
\newcommand{\Emin}{\forcemath{E_{\rm min}}}

\ifx \degree \undefined
  \newcommand{\degree}{\ifmmode {^{\circ}} \else {$^{\circ}$} \fi}
\fi
\newcommand{\degrees}{\ifmmode {^{\circ}} \else {$^{\circ}$} \fi}

\newcommand{\quarter}{\ifmmode {\frac{1}{4}} \else {$\frac{1}{4}$} \fi}

\ifx \km \undefined
  \newcommand{\cm}{\unit{cm}}
\else
  \renewcommand{\cm}{\unit{\cm}}
\fi
\newcommand{\cmpersecsqr}{\unit{\cm\,\s^{-2}}}

\newcommand{\wattpermsqr}{\unit{W\ m^{-2}}}
\newcommand{\joulepermsqr}{\unit{J\ m^{-2}}}
\newcommand{\hzpersec}{\unit{Hz\ s^{-1}}}
\newcommand{\chisqr}{\unitnospace{\chi^2}}
\ifx \km \undefined
  \newcommand{\km}{\unit{km}}
\else 
  \renewcommand{\km}{\unit{km}}
\fi

\ifx \s \undefined
  \newcommand{\s}{\unit{s}}
\else
  \renewcommand{\s}{\unit{s}}
\fi

\newcommand{\persec}{\unit{s^{-1}}}

\ifx \us \undefined
  \newcommand{\us}{\unit{\mu s}}
\else
  \renewcommand{\us}{\unit{\mu s}}
\fi

\ifx \Hz \undefined
  \newcommand{\Hz}{\unit{Hz}}
\else 
  \renewcommand{\Hz}{\unit{Hz}}
\fi
\ifx \kHz \undefined
  \newcommand{\kHz}{\unit{kHz}}
\else
  \renewcommand{\kHz}{\unit{kHz}}
\fi
\ifx \MHz \undefined
  \newcommand{\MHz}{\unit{MHz}}
\else
  \renewcommand{\MHz}{\unit{MHz}}
\fi
\ifx \GHz \undefined
  \newcommand{\GHz}{\unit{GHz}}
\else
  \renewcommand{\GHz}{\unit{GHz}}
\fi

\newcommand{\lt}{\unit{<}}
\newcommand{\gt}{\unit{>}}

\newcommand{\tten}[1]{\ifmmode {\times 10^{#1}} \else {$\times 10^{#1}$} \fi}
\newcommand{\tentothe}[1]{\ifmmode {10^{#1}} \else {$10^{#1}$} \fi}

\newcommand{\kpc}{\unit{kpc}}

\newcommand{\DM}{\unit{DM}}
\ifx \dm \undefined
  \newcommand{\dm}{\unit{pc\ cm^{-3}}}
\fi

\newcommand{\HI}{{\rm H{\sc i}}}

\newcommand{\doublet}{\ifmmode {\lambda\lambda} \else {$\lambda\lambda$} \fi}
\newcommand{\singlet}{\ifmmode {\lambda} \else {$\lambda$} \fi}

\newcommand{\kmpersec}{\km\persec}

\newcommand{\percmcubed}{\unit{cm^{-3}}}

\newcommand{\Msps}{\unit{{\rm Msps}}}
\ifx \jahh \undefined
  \newcommand{\jahh}{{\rm JAHH}}
\fi
\ifx \jgr \undefined 
  \newcommand{\jgr}{{J.\,Geophys.\,Res.\xspace}}
\fi
\ifx \pasp \undefined
  \newcommand{\pasp}{{Pub.\,Astron.\,Soc.\,Pac.\xspace}}
\fi
\ifx \jgrsp \undefined
  \newcommand{\jgrsp}{{J.\,Geophys.\,Res.\,(Space\,Phys.)\xspace}}
\fi
\ifx \aspc \undefined
  \newcommand{\aspc}{{ASP\,Conf.\,Ser.\xspace}}
\fi
\ifx \iauc \undefined
  \newcommand{\iauc}{{IAU~Colloq.}}
\fi
\ifx \grl \undefined
  \newcommand{\grl}{{Geophys.\,Res.\,Let.\xspace}}
\fi
\ifx \solphys \undefined
\newcommand{\solphys}{{Sol.\,Phys.\xspace}}
\fi
\ifx \aj \undefined
  \newcommand{\aj}{{Astron.\,J.\xspace}}
\fi
\ifx \apj \undefined
  \newcommand{\apj}{{Astrophys.\,J.\xspace}}
\fi
\ifx \apjl \undefined
  \newcommand{\apjl}{{Astrophys.\,J.\xspace}}
\fi
\ifx \apjs \undefined
  \newcommand{\apjs}{{Astrophys.\,J.\,Suppl.\xspace}}
\fi
\ifx \aap \undefined
  \newcommand{\aap}{{Astron.\,&\,Astrophys.\xspace}}
\fi
\ifx \aas \undefined
  \newcommand{\aas}{{AAS Meeting Abstracts\xspace}}
\fi
\ifx \procspie \undefined 
  \newcommand{\procspie}{{Proc.\,SPIE}}
\fi
\ifx \spiec \undefined
  \newcommand{\spiec}{\procspie}
\fi
\ifx \ieeeproc \undefined
  
\fi
\ifx \ieeeproc \undefined
  \newcommand{\ieeeproc}{IEEE\,Proc.\xspace}
\fi
\ifx \jgridcomp \undefined
  \newcommand{\jgridcomp}{{J.\,Grid\,Comput.\xspace}}
\fi
\ifx \revsciinst \undefined
  \newcommand{\revsciinst}{{Rev.\,Sci.\,Instr.\xspace}}
\fi
\ifx \skytel \undefined
  \newcommand{\skytel}{{Sky\,\&\,Telescope}}
\fi
\ifx \jai \undefined
  \newcommand{\jai}{{\rm J. Astron. Instr.}}
\fi
\ifx \areps \undefined
  \newcommand{\areps}{{\rm Annu. Rev. Earth \& Planet. Sci.}}
\fi
\ifx \detokenize \undefined
  \newcommand{\detokenize}[1]{{#1}}
\fi
\ifx \doi \undefined
  \newcommand{\doi}[1]{DOI \href{http://dx.doi.org/\detokenize{#1}}{\ttfamily \detokenize{#1}}\spc}
\fi
\ifx \cise \undefined
  \newcommand{\cise}{{Computi. Sci. \& Eng.}}
\fi

\newcommand{\var}[1]{{\ifmmode {{#1}}\else {${#1}$}\fi}\xspace}

\newcommand{\nchirps}{\var{123\,000}\xspace}
\newcommand{\minbin}{\var{0.075\,\Hz}}
\newcommand{\mintimebin}{\var{8.1\tten{-3}\,\s}}
\newcommand{\maxbin}{\var{1221\,\Hz}}
\newcommand{\maxtimebin}{\var{13.4\,\s}}
\newcommand{\minchirp}{\var{-100\,\hzpersec}}
\newcommand{\maxchirp}{\var{100\,\hzpersec}}

\newcommand{\fullband}{\var{2.5\,\MHz}}
\newcommand{\recordersamplerate}{\var{2.5\,\Msps}}
\newcommand{\bandcenter}{\var{1.42\,\GHz}}

\newcommand{\chirprange}{\var{\pm 100\,\hzpersec}}

\newcommand{\wubandwidth}{\var{9.766\,\kHz}}
\newcommand{\wusamples}{\var{2^{20}}}
\newcommand{\wuduration}{\var{107.37\,\s}}

\begin{document}

\title[SETI@home I: Data Acquisition and Front-End Processing]{SETI@home: Data Acquisition and Front-End Processing}
\thanks{https://setiathome.berkeley.edu/}

\author{E.~J.~Korpela}
\affiliation{ 
Space Sciences Laboratory, University of California, Berkeley, CA 94720
}
\email{korpela@berkeley.edu}
 
\author{D.~P.~Anderson}
\affiliation{ 
Space Sciences Laboratory, University of California, Berkeley, CA 94720
}
\email{davea@berkeley.edu}
\author{J.~Cobb}
\affiliation{ 
Space Sciences Laboratory, University of California, Berkeley, CA 94720
}
\email{jeffcobb@berkeley.edu}
\author{M.~Lebofsky}
\affiliation{ 
Space Sciences Laboratory, University of California, Berkeley, CA 94720
}
\email{lebofsky@berkeley.edu}
\author{W.~Liu}
\affiliation{ 
Space Sciences Laboratory, University of California, Berkeley, CA 94720
}
\email{liuwei_berkeley@berkeley.edu}
\author{D.~Werthimer}
\affiliation{ 
Space Sciences Laboratory, University of California, Berkeley, CA 94720
}
\email{danseti@berkeley.edu}

\affiliation{ 
Space Sciences Laboratory, University of California, Berkeley, CA 94720
}

\date{\today}

\begin{abstract}
SETI@home is a radio Search for {Extraterrestrial} Intelligence (SETI) project, looking for technosignatures in data recorded
at {multiple observatories} from 1998 to {2020}.
Most radio SETI projects analyze data using
dedicated processing hardware.
SETI@home uses a different approach:
time-domain data is distributed over the Internet to $>10^5$ volunteered home computers, which {analyze} it.
The large amount of computing power {this} affords
($\sim10^{15}$ {floating-point} operations per second {(FPOP/s)})
allows us to increase the sensitivity and generality of our search in three ways.
We use coherent integration, a technique in which data is
transformed so that the power of drifting signals is confined to a single discrete Fourier transform (DFT) bin.
We perform this coherent search over \nchirps Doppler drift rates
in the range ($\chirprange$).
Second, we search for a variety of signal types,
such as pulsed signals
and arbitrary repeated waveforms.
The analysis uses a range of DFT sizes, with frequency resolutions
ranging from \minbin to \maxbin.
The {front end} of SETI@home produces
a set of {\em detections}
that exceed thresholds in power and goodness of fit.
We accumulated $\sim1.2\tten{10}$ such detections.
The {back end} of SETI@home takes these detections,
identifies and removes radio frequency interference (RFI),
and looks for groups of detections that are consistent with extraterrestrial origin
and that persist over long timescales.
This paper describes the {front end} of SETI@home {and provides parameters for the 
primary data source, the Arecibo Observatory;
the back end} and its results are described in a companion paper.
\end{abstract}


\section{\label{sec:intro}Introduction}
{
\subsection{Background}

The question of whether life exists in other parts of the universe is
important and unanswered.
The 1952 Muller-Urey experiment \citep{miller1953,miller1959}
demonstrated the possibility of abiotic
production of the molecular components of living systems.
The detection of
amino acids in meteorites \citep{pearce2015}
and prebiotic molecules in interstellar space \citep{zeng2019, rivilla2023} showed that such
processes are possible
even outside a planetary atmosphere.

The direct detection of living organisms
outside the Solar System remains unlikely in the near future.
A more likely scenario is an indirect detection,
such as an atmospheric biosignature: a compound
released into the atmosphere by biological processes.
However, such compounds may also have an abiogenic source,
so whether such a detection indicates life is uncertain \citep{tokadjian24,court12}.

Detection of intelligence would provide more certain evidence of life.
An extraterrestrial intelligence (ETI) could create artifacts, signals, or processes that are detectable at interstellar
distances and have no natural counterpart.
Such processes could be a form of radiation
(electromagnetic, particle, or gravitational)
or a physical artifact
(a spacecraft or object passing through or remaining in the Solar System, 
a structure detectable at interstellar distance, 
or an atmospheric component that only has
a technological means of production).
These are collectively known as
{\em technosignatures} \citep{technosignatures}.

Due to the relative ease of creating and detecting radio waves
and the relative transparency of atmospheres and interstellar space to such waves,
radio has been proposed as a means of detecting extraterrestrial
intelligence \citep{cocconi59}.
Two primary approaches have been used for such searches:
{\em sky surveys} cover a large fraction of the solid angle of the entire sky, and {\em targeted searches} focus on individual stars or galaxies \citep{search_strategy}.  Such searches have been collectively known as the
Search for Extraterrestrial Intelligence (SETI).

Several targeted searches have been performed,
including OZMA and OZMA II at Green Bank
\citep{drake60, sagan75, drake86, gray21}
, Phoenix at the Arecibo Observatory
\citep{backus02}
and at the Allen Telescope Array (ATA),
and Breakthrough Listen projects at the Parkes and Green Bank observatories
\citep{price20, enriquez17}. 
Recently, Breakthrough Listen has begun to observe targets at the Very Large Array \citep{tremblay24} and MeerKAT \citep{czech21}.
In addition, observations of multiple targets have been made at
the FAST observatory in China \citep{luan23} and the ATA \citep{tusay24}.

There have also been a number of sky surveys.
Some have operated {\em commensally},
collecting data from a telescope while
its pointing was being controlled by other projects.
These include searches using various
generations of the SERENDIP spectrometer
at the Hat Creek and Green Bank observatories
\citep{werthimer88}
and at the Arecibo observatory
\citep{cobb00, bowyer16}.
Other sky surveys used dedicated telescopes.
These include the early Ohio State
project and its ``Wow!" signal
\citep{kraus77}, the ``Fly's Eye" project \citep{siemion12} and a brief survey of the anti-solar point \citep{hort24}
at the ATA.  

To date, no repeatable detections of interstellar
technosignatures have been made.

Because there are no known sources of narrowband emissions, radio SETI searches have typically searched for narrowband signals.
The frequency range and the number and width of channels have been limited by available technology.
The first searches used existing instruments with channel widths from 100 Hz to tens of kHz \citep{drake60,kraus77,bowyer80}.
As technology progressed, special purpose SETI spectrometers were developed that used Fourier transform processors,
programmable gate arrays (PGAs) and graphics processing units (GPUs) \citep{werthimer95, siemion11, archer16}.
The frequency range of these spectrometers increased from kHz to GHz, while channel widths decreased to $\sim 1\,\Hz$, enlarging search space coverage and improving sensitivity.
As will be discussed in \S\ref{sec:doppler}, further reduction in channel bandwidth requires correction for Doppler effects.
}

\subsection{SETI@home}
\begin{figure}[tbp]
\centerline{
\includegraphics[width=\columnwidth]{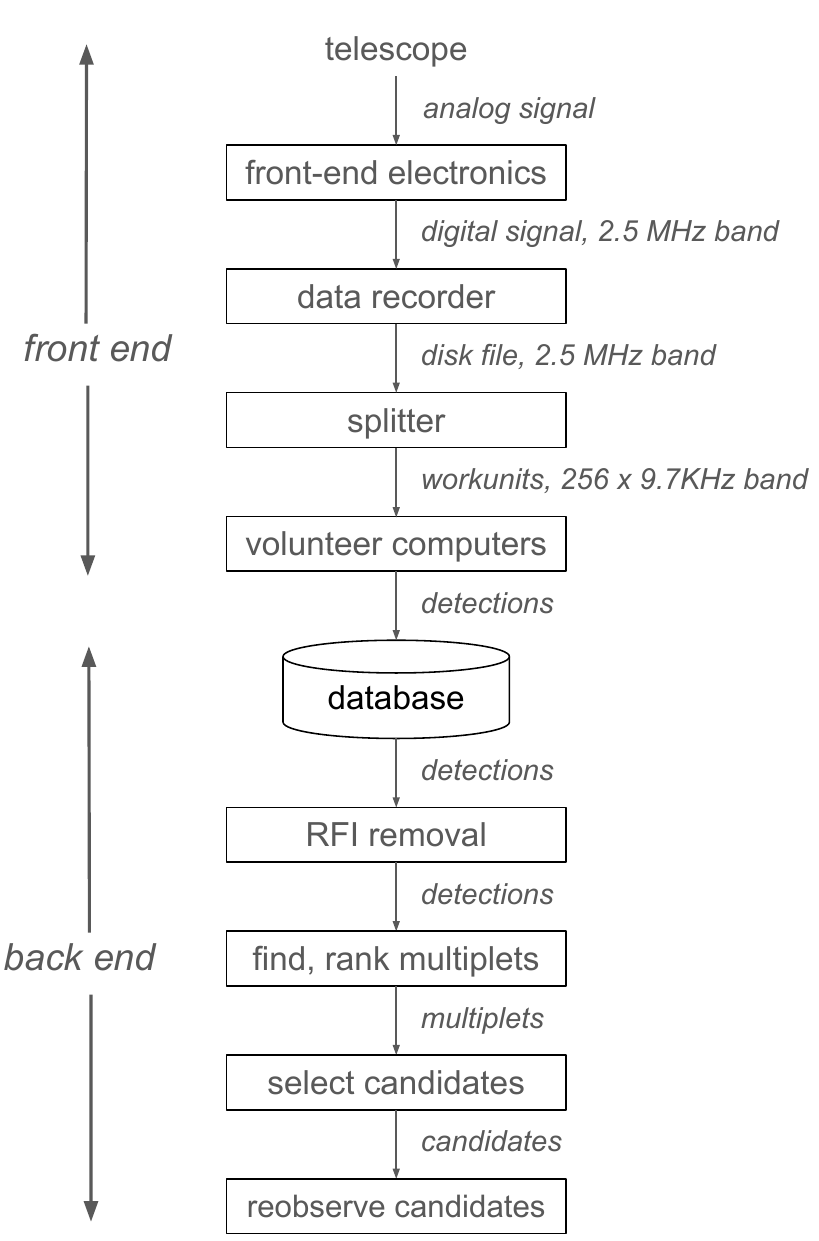}
}
\caption{The SETI@home data {acquisition and }analysis {pipeline.}}
\label{fig:pipeline}
\end{figure}

SETI@home is a radio Search for Extraterrestrial Intelligence (SETI) project{, which searched for}
for several types of signals in {recorded data.
Most of this data was
recorded} commensally at the Arecibo observatory over a 22-year period.
{Other data from the Parkes and Green Bank observatories was provided by Breakthrough Listen \citep{breakthroughlisten}.}
The first stage of data analysis finds {\em detections}:
brief and statistically unlikely excesses of {continuous or pulsed
narrowband} power.
The second stage, described in \citet{results_paper}, removes RFI and identifies and ranks {the} target signal candidates
(see Figure \ref{fig:pipeline}).

Most radio SETI projects process data in near real-time using special purpose analyzers at the telescope.
SETI@home takes a different approach.
It records digital time-domain (also called baseband) data, and distributes it over the Internet
to large numbers of computers that process the data,
using both CPUs and GPUs.

This approach requires recording and storing
an amount of data proportional to the frequency range we cover,
and transmitting this data through home Internet connections.
These factors impose performance constraints that allow us to examine only
a relatively narrow frequency range \mbox{(\fullband)}.

However, the approach provides a large amount of computing power
(roughly $\sim10^{15}$ floating point operations per second)
with which to analyze this data.
We use this in several ways
to increase the sensitivity and generality of our search.
First, we use coherent integration, a technique in which data is
transformed so that the power of drifting signals is confined to a single discrete Fourier transform (DFT) bin.
We perform this coherent search over \nchirps Doppler drift rates
in the range $\chirprange$.
Second, we search for a variety of signal types,
such as pulsed signals (using a fast-folding algorithm)
and repeated non-sinusoidal waveforms {through autocorrelation} of the time-domain data).

Third, the analysis uses a range of 15 DFT sizes, with frequency resolutions
ranging from \minbin (\maxtimebin) to \maxbin (\mintimebin).
{The longest DFT length was chosen to be the best power-of-two match to the most common observation length (13.7\,s, or one beamwidth at sidereal rate), while the shortest was chosen to be the widest power-of-two frequency bin that would not be affected by removal of natural \HI\ line emission;
see \S\ref{sec:algo}.

The parameters of SETI@home at Arecibo are summarized in Table \ref{table:params}.  Parameters for observations made at other observatories will be provided in future publications related to
those observations.
}

\begin{table*}[tbp]
\begin{center}
{
\begin{tabular}{l l}
Parameter & Value \\
\hline
Frequency range & 2.5 MHz centered at 1.42 GHz \\
Sample size & 2 bit complex recording, 4 bit complex distribution \\
Doppler correction & $\pm$100 Hz/s, coherent \\
Signal types detected & narrowband (continuous and pulsed), autocorrelation \\
Number of frequency & 15 (0.075 Hz to 1221 Hz) \\ resolutions &  \\
Observatory & Arecibo \\
Receivers: & L-band flat feed (1-beam, single polarization)\\
 & ALFA (7-beam, dual polarization)\\
Angular resolution & 2.5$^\prime$\\
Sky coverage & 12\,375 square degrees (30\% of the celestial sphere) \\
System Temperature & 25-29K~typ. \\
Effective Area & 10\,900 m$^{2}$ \\
Observation period: & \\
 -- L-band flat feed & 1999-2006, 386 days total \\
 -- ALFA & 2006-2020, 400 days total \\
\end{tabular}
}
\caption{{Parameters of SETI@home at Arecibo\label{table:params}}}
\end{center}
\end{table*}

This paper is organized as follows.
\S\ref{sec:target} describes the types and range of target signals {that} SETI@home is looking for.
\S\ref{sec:data} discusses the way we record, store, and distribute time-domain data.
\S\ref{sec:detections} {describes} the types of detections
{(momentary signals)} that we look for,
how we find them, and how we assign scores to them.
\S\ref{sec:analysis} describes the data analysis algorithm
and the sensitivity it achieves.
{\S\ref{sec:testing} describes how we tested the application using both simulated and real data.}
\S\ref{sec:volunteer} discusses the use of volunteer computing.
\S\ref{sec:related} compares SETI@home to related projects,
and \S\ref{sec:conclusion} gives conclusions and discusses possible future work.

{SETI@home is licensed under the General Public License (GPL).}  The SETI@home source code, written in C++, is available at
\href{https://sourceforge.net/projects/seti-boinc/}
{https://sourceforge.net/projects/seti-boinc/}.   {The source code for the simulated data generator described in \S\ref{sec:testing} is located in the tools subdirectory of this repository.  The source code for the software radar blanker is available at \href{https://sourceforge.net/projects/seti-science/}{https://sourceforge.net/projects/seti-science/} in the "software\_blanking" subdirectory.}
SETI@home {also} uses the ``setilib" library available at \href{https://sourceforge.net/projects/setilib/}
{https://sourceforge.net/projects/setilib/}.

\section{\label{sec:target}Target signals}

SETI@home looks for a range of target signals - 
signals with characteristics consistent with technological origin that are not
known to occur naturally.
Specifically, we look for:
\begin{itemize}
\item Continuous signals with $\Delta\nu \leq  1221 ~ \Hz$; that is, signals whose bandwidth is small enough that most of the power is
concentrated in a single DFT bin.
\item Periodically pulsed {narrowband} signals,
which turn on and off with some period, pulse duration, and phase, on timescales small compared to the sidereal rate beam crossing time (13 seconds),
and with $\Delta\nu\Delta t\sim1$ where $\Delta t$ is the pulse duration.
\item Arbitrary waveforms that repeat after {a} short delay.
\end{itemize}

We look for such signals occurring either
transiently (for a few seconds or minutes)
or persistently
(over a long period, potentially the entire observation period).

We assume that the signal transmitter is either
1) in an inertial frame at nearly constant velocity relative to the Galactic barycenter, 2) on the surface of a rotating planet orbiting a star, 3) in orbit around a planet orbiting a star, or 4) directly in orbit around a star.
The front-end analysis could, in principle, be used to search for objects within the Solar System, but this is not done by the existing back-end  \citep{results_paper}.

\subsection{\label{sec:doppler}Doppler drift}

The frequency at which {a telescope} receives a signal
is Doppler-shifted by the velocity components of both
transmitter and receiver in the direction of the signal path. {In this section, we use
specific values related to data collected using the SETI@home data recorder at Arecibo, where the majority of the observations analyzed by SETI@home were conducted.}
The shift due to receiver motion, and the time derivative of the shift, are known;
they correspond to the various accelerations of
the receiver's reference frame,
due primarily to the Earth's rotation and orbit.
We can apply corrections that shift a received frequency to the frequency that would be observed at the Galactic barycenter,
which can be considered an inertial frame over the times scales of the observations; see \citet{doppler_code_source} and references therein.
The maximum velocity difference between the receiver and this frame is $\pm$29.9 \kmpersec,
corresponding to a frequency shift of $\pm$142 kHz at \bandcenter.

However, we don't know if the transmitter
is applying a similar correction to the transmission frequency.
If a signal is intended as a beacon and is directional,
it could be corrected for the accelerations of the transmitter to
present a stable frequency for the observer.
We refer to such corrected signals as {\em barycentric} because they will be at a constant
frequency in the frame of the barycenter of the solar system.
After the receiver Doppler correction is applied,
these signals will appear at a nearly constant frequency.

Because the correction applied to a transmitted signal {depends} on the
direction of the receiver, leakage signals or omnidirectional beacons
are unlikely to be corrected in this manner. 
Such signals would have a Doppler shift corresponding
to the radial velocity of the transmitter.
We call such signals {\em non-barycentric} as they would not be frequency stable in
the barycentric frame.
After correction for the receiver Doppler shift,
they will still appear to be varying in frequency.

The {ranges} of the sender Doppler shift
and its derivative (Doppler drift rate)
depend on the movements of the transmitter.
We look for target signals for which these
ranges are consistent with certain assumptions
about the movements of the transmitter, such as the rotational rate of
planetary transmitters; see \S\ref{sec:coherent}. 

\subsection{\label{sec:dispersion}Interstellar Dispersion}

During propagation through the interstellar medium,
signals of nonzero bandwidth become dispersed due to interaction with free electrons.
The amount of this dispersion depends upon the amount of ionized material through which the signal propagates.
The total differential delay between the lowest and highest frequency {components} of the signal is
\begin{equation}
\Delta t = (8.3 \us) \frac{\Delta\nu [\MHz]}{\nu^{3} [\GHz]} \DM [\dm]
\end{equation}
where \DM\ is the dispersion measure, defined as the electron column density between the transmitter and receiver:
$\DM\equiv \int_0^D n_e dl$, expressed in units of \dm \citep{lorimer12}.
As the typical interstellar electron density in the Galactic plane is $n_e \sim 0.08\,\percmcubed$ \citep{taylor93}, Galactic \DM\ is $DM_G\simlt 8 D[\kpc]$ where $D[\kpc]$ is the transmitter-receiver distance in \kpc,
therefore a 10 \kpc\ range would lead to a $\langle\DM\rangle \simlt 800\,\dm$.
The median \DM of known Galactic pulsars is $\sim 140\,\dm$, {(\citeauthor{atnf_pulsar} \citeyear{atnf_pulsar}, \href{https://www.atnf.csiro.au/research/pulsar/psrcat}{https://www.atnf.csiro.au/research/pulsar/psrcat})} which is likely due to the scale height of pulsars placing them above or below the Galactic plane,
and to a detection bias selecting for nearby pulsars.  
To prevent receivers from needing to correct for dispersion for a DM of 800\,\dm,
an extraterrestrial intelligence (ETI) might choose to send a $\Delta \nu\Delta t\sim 1$ beacon with bandwidth
\begin{equation}    \Delta \nu \ll \left( \frac{\nu^3 [\GHz]}{8.3\tten{-12} \langle\DM\rangle} \right)^\frac{1}{2} \Hz \sim 21\,\kHz\ {\rm at\ 1.42\ GHz.}
\end{equation}
Because SETI@home only considers signal bandwidths $\Delta\nu\leq\maxbin$,
dispersion is unimportant to the analysis.
The limiting dispersion measure for a signal at 1.42 \GHz with \maxbin\ bandwidth is $\DM\lt2.3\tten{5}\ \dm$,
well above the \DM of any known Galactic pulsar.

To study the case where dispersion is important, we operated a sister project, Astropulse, 
using the same data source and volunteer computing infrastructure as SETI@home.
Astropulse looked for
single and repeated broadband pulses,
with many possible origins including both technosignatures and
astrophysical phenomena such as black hole evaporation and pulsars.
Astropulse workunits included the full 2.5 MHz band,
and it looked for pulses in the dispersion measure range $49.5\,\dm \leq |DM| \leq 830\,\dm$ using coherent de-dispersion.
Astropulse is described in \citet{vonkorff13}.

\section{\label{sec:data}Data acquisition and initial processing}

\subsection{\label{sec:ao_obs}Arecibo observations}

Before 2006, SETI@home obtained data from an L-band flat feed
mounted on a carriage house opposite the Arecibo Gregorian reflector dome.

After the 2006 installation of the 7-beam Arecibo L-band feed array (ALFA),
SETI@home used ALFA as its data source.
ALFA is an array of seven receivers arranged in a hexagonal pattern with one in the middle, which was mounted in the Gregorian dome.
SETI@home made its observations commensally, in conjunction with other uses of the ALFA array.
Over the course of the project, the array was used to search for pulsars near the plane of the Galaxy, to map the distribution of hydrogen in all parts of the Galaxy visible from Arecibo, and to search for extragalactic hydrogen gas in isolated clouds or in nearby galaxies.
This resulted in three main modes of observation.
The pulsar surveys tended to track positions in the sky for 30 seconds to tens of minutes while accumulating data.
The other {surveys used} {either} a drift scan mode{, in which} the receivers are held in position while objects in the sky drifted through telescope beams due to the earth's rotation,
or a ``basket-weave'' mode in which the receiver tracked north and south while the sky drifted by, resulting in a zigzag path {\citep{peek11}}.  

If the primary feed was stationary, objects in the sky passed through the beam of one of the ALFA receivers (0.05\degrees) at the sidereal rate.
An object would require $\sim$13 seconds to transit the field.
When used in basket-weave mode, less time was required for transit.
When tracking, objects could remain in the field of view for a long duration,
up to a possible maximum of $\sim4$ hours.

Using the ALFA receiver, the telescope could view
declinations between -2\degrees and 38\degrees,
or about 25\% of the sky. 
Our observations covered almost this entire area, most of it multiple times;
see \cite{results_paper}.

The SETI@home data recorder recorded a \fullband band from each of the two linear polarizations of the seven receivers (14 data streams in all) centered at \bandcenter near the \HI\ hyperfine transition at 1.4204 GHz.
We chose the hydrogen line because it is considered to be a likely frequency for deliberate transmissions.
Extraterrestrial astronomers who are aware of the \HI\ transition are likely to use it to survey the structure of the galaxy.
The potentially large number of observers makes this frequency a good choice for transmissions designed to attract attention.

The 14 analog signals from ALFA'’s 7 receivers were simultaneously fed into several different instruments including spectrometers,
pulsar and fast radio burst search machines, as well as the SETI@home data recorder.
This allowed several different experiments to observe the sky simultaneously.
{The front end electronics} down-converted {the analog signal,} extracted the \fullband band centered at \bandcenter and conver{ted} the signals to complex baseband.
Each of the 14 complex baseband signals {was recorded} at \recordersamplerate,
with each sample being a 2-bit complex number (one bit real and one bit imaginary){, along with the observatory radar blanking signal}.
The data was recorded continuously onto hot-swappable disk drives.
The disk drives were physically shipped to Berkeley for analysis.
{This} raw data (about 1 petabyte) is archived at the National Energy Research Scientific Computing Center at the Lawrence Berkeley Laboratory.  

\subsubsection{\label{sec:radar}Radar Blanking}

There are several strong radars on the island of Puerto Rico. SETI@home employed both software and hardware to mitigate interference from these radars. When radars were contaminating the SETI@home data, we replaced the time domain data from the receivers with shaped random noise.  Because this interference was from low duty cycle (pulsed) radars, the sensitivity loss from radar blanking was low, about 2.5\%.     

SETI@home used two radar mitigation strategie{s at Arecibo}.
The first, {\em hardware radar blanking},
{used} a small dedicated antenna and receiver system designed by the observatory to detect radar signals \citep{perillatwebsite}.
The digital radar on/off signal output from this system was recorded by the SETI@home data recorder along with the time domain science data from the telescope's receivers.
In post-processing, when the pulsed radar is on, shaped random noise matching the frequency sensitivity of the receiver to noise is substituted for data from the receiver.        
 
The second strategy, {\em software radar blanking},
searches for radar interference in the receiver data by cross-correlating the time domain data with five different known radar patterns detected at Arecibo.
If the correlation is above a threshold for any of these five templates,
the receiver data during the expected radar pulses is replaced with shaped noise.

{Data obtained from other observatories was typically not radar blanked by the SETI@home front end.}

\subsection{\label{sec:splitter}Data Splitting}
A {\em splitter} program divides the data {from each receiver polarization channel} into {256} frequency {subbands} of about \wubandwidth each and lengths of \wusamples samples (\wuduration in duration).
We call these segments {\em workunits}.
Early versions of the splitter used {2048-point forward / 8-point inverse} DFT filtering to break the band up into 256 sharply defined {subbands}.
Later versions used a polyphase filter bank to improve 
{out-of-band} rejection.  
Originally, the workunits were resampled to 2-bit complex for compactness,
but as typical internet bandwidth increased, this was changed to 4-bit complex samples in order to reduce quantization losses. \citep{quantization}

Sequential workunits of a given {subband} are overlapped in time by approximately 20 seconds so that the typical longest features of interest -- 13 seconds or so -- are always contained entirely {within} at least one workunit.

{Each workunit included a data header containing all of the parameters used by the SETI@home client application.  This includes the time as Julian date, the parameters of the telescope (name, astronomical and geodetic location), the receiver system (frequency, bandwidth), the splitting method (workunit bandwidth, number of samples per workunit, center frequency, and other parameters necessary to determine the frequency of a signal within a workunit), and celestial coordinates of the beam center throughout the duration of the workunit.

The data itself could be output in any of the encodings and bit-widths supported by the SETI@home application.  Complex samples with power-of-two sizes from 2-bits to 16-bits were supported by default.  Typically SETI@home used a base64-like encoding, but SETI@home also supports binary, multiple XML encoding forms, base64, base85, CSV, quoted-printable, and hexidecimal, in addition to ASCII floating point.

\subsection{\label{sec:other}Observations at other observatories}

SETI@home was designed to be agnostic to the source of the data to the extent possible.  Over
the course of the project, baseband data was also collected by the Breakthrough Listen project at both the 64-meter Parkes Telescope and 100-meter Green Bank Telescope (GBT) and analyzed by SETI@home.  Tests of data from LOFAR and the 25-meter Dwingeloo telescope were also conducted,
but this data was not widely distributed and the results were not inserted into the SETI@home
database.  Reobservations of candidates are being conducted at the FAST observatory, and this data will be analyzed using the SETI@home client. Some amateur radio astronomers have extended SETI@home to common recording formats including lossless audio formats such as .WAV or .FLAC and
binary formats used by GNUradio. Such extensions are not included in the official repository.
}

\section{\label{sec:detections}Detections}

Each workunit is analyzed by a program called the SETI@home client.
The client looks for {\em detections}, which are artifacts or possible signals.
There are five types of detection: {\em spikes}, {\em Gaussians}, {\em pulses}, {\em triplets},
and {\em autocorrelations}.
{Table \ref{table:detection_types} gives a brief description of each type, and for observations conducted at Arecibo, its primary parameter range, its typical sensitivity, and the range at which a transmitter with average equivalent isotropic radiated power (EIRP) of 20 TW (similar to the Arecibo planetary radar EIRP) could be detected.}
These types span our range of target signals (see \S\ref{sec:target}).

\begin{table*}[tbp]
\begin{center}
\begin{tabular}{l l c c c}
\multicolumn{1}{c}{Type} & \multicolumn{1}{c}{Signal} & \multicolumn{1}{c}{Parameter} & \multicolumn{1}{c}{Event Sensitivity} & \multicolumn{1}{c}{20 TW EIRP} \\
  & \multicolumn{1}{c}{description} & \multicolumn{1}{c}{Range} & \multicolumn{1}{c}{typ.} & \multicolumn{1}{c}{detection distance} \\
 & & & \multicolumn{1}{c}{[\wattpermsqr]} & [pc] \\
\hline
Spike & Continuous narrowband & $0.074\,\Hz \leq \Delta\nu \leq 1220\,\Hz$ & $1.4\tten{-25} $ & 110 \\
Gaussian & Continuous narrowband & $0.60\,\Hz \leq \Delta\nu \leq 1220\,\Hz$&$1.1\tten{-25} $ &123 \\
Pulse & Pulsed narrowband & $1.6\,{\rm ms}\leq p \leq 35.79\,{\rm s}$ & $1.2\tten{-25}$ & 118\\
Triplet & Pulsed narrowband &  $4.2\,{\rm ms}\leq p \leq  53.69 {\rm s}$ & $7.9\tten{-26}$ & 145 \\
Autocorrelation & Any repeated waveform &  $|\tau|\leq 6.7s$ & $1.5\tten{-25}$ & 106\\
\end{tabular}
\caption{{Detection types, their parameter ranges, and their sensitivity for SETI@home observations made at Arecibo.\label{table:detection_types}}}
\end{center}
\end{table*}

{Each detection} has parameters
(power and, for Gaussians, goodness of fit)
that reflect its significance.
The algorithm for each type has thresholds for these parameters.
{The thresholds for each type} are chosen so that the number of
{\em false alarms} ({above-threshold detections} in 
data consisting of random Gaussian noise) per workunit is about one.

{The client returns a detection if its parameters exceed the thresholds.
For a given workunit, the client also returns the ``best'' detection of each type even if it does not exceed the thresholds.
This allows proper operation of the client to be checked even if no detections
are above threshold.
}

When a detection $D$  is returned, it is assigned a {\em probability score}, $S(D)$,
proportional to the {an estimate of the} probability of that detection resulting from random noise. 

In the following sections, we describe the algorithm for each detection type,
including how its thresholds and probability scores are computed.
The algorithms operate on frequency-domain data
computed as follows (see \S\ref{sec:analysis}).
The client uses coherent integration at a wide range of Doppler drift rates
and uses a range of channel widths (or DFT lengths).
At each combination of Doppler drift rate and DFT length,
the client computes a sequence of DFTs on the de-drifted time-domain data.
This produces, for each frequency channel,
a power-versus-time (PvT) array.

\subsection{\label{sec:spike}Spikes}
Each spectrum generated by the DFT is examined for bins with power above a threshold.
This threshold is 24 times the mean power in that spectrum, which, given complex data with a random Gaussian distribution,
results in an {$e^{-24}$} probability that a single bin exceeds the threshold.\footnote{These false alarm probability estimates are based on the unrealistic assumption of evenly sampled data, an infinite time series, and the presence of no signals apart from non-truncated Gaussian noise \citep{percival93,baluev08}.  They are not expected to be fully accurate estimates of the false alarm probability and become increasingly inaccurate in the presence of short DFT lengths, multiple and/or strong signals, and signal truncation.  The calculated thresholds are used only to identify signals for further processing and as a relative comparison of the statistical improbability of signals.}  This threshold was chosen because it usually results in $\sim$ a few detections in each workunit of actual data.
Detections above this threshold, including their parameters such as position, frequency, channel bandwidth, and Doppler drift rate are returned by the client.

The power in an individual spectral bin is a magnitude of a complex number,
so the power distribution per bin can be represented as $\chi^2$ distribution with 2 degrees of freedom (DOF).
Hence we define the detection probability score for a spike $D$ as
\begin{equation}
{S_{\rm spike}(D)=Q(\chi^2|2)=Q(1,\frac{\chi^2}{2})=\frac{1}{\Gamma(1)}\int_{\frac{\chi^2}{2}}^\infty e^{-t} dt = e^{-\frac{P(D)}{\langle P\rangle }}}
\end{equation}
where $Q$ is the complementary incomplete gamma function,
$P(D)$ is the spike power and $\langle P\rangle $ is the frequency averaged power in the DFT.

\subsection{\label{sec:gaussian}Gaussians}
If the telescope beam is moving with sufficient speed across the sky,
a signal would be visible in that beam for less than the duration of the workunit.
As the beam passes over the signal, its detected power in a PvT
array would match the sensitivity profile of the beam, which for constant motion is nearly Gaussian in shape.  

The client performs Gaussian fitting on each PvT array if the time resolution is sufficient ($\frac{L_{\rm data}}{L_{\rm DFT}} \geq 64$), and if the angle traversed, $\theta$, is sufficient for both the Gaussian shape and a background level to be determined (4.5 beam widths $\lt  \theta \lt $ 22.5 beam widths). Because the rate of motion is known, the $\sigma$ width of the Gaussian is a known quantity, {but it varies between workunits depending on the rate of telescope motion.
If the telescope beam is moving too slowly across the sky ($\lt 4.5$ beamwidths per workunit) or too rapidly ($\gt 22.5$ beamwidths
per workunit), Gaussian fitting is not performed.}

The client rebins the PvT array for the channel by co-adding adjacent time bins to obtain a 64 element array which is used for the subsequent step.   {A 64-point array was chosen to limit the size of the array used for the Gaussian fit, resulting in faster run times.  It also led to a similar analysis regardless of DFT length, reducing the complexity of post-processing.}

First, this rebinned array is searched to see if there is a bin with power {greater than} 3 times the mean power of the array.
If not, the search for this array is abandoned. 

The client then loops through each of the 64 elements of the array,
presuming the peak location to be at that element{,} and determin{es} the background level using the array elements that are farther than $2\sigma$ from the peak location.
The best-fit peak amplitude is determined {for} each element.
The client determines the reduced $\chi^2$ value for this 2 parameter (peak level, peak position, DOF=62) and a non-Gaussian invariant background (1 parameter, DOF=63).
The threshold conditions are reduced \chisqr\ of the Gaussian fit $\chisqr_{\rm red}<1.42\ (\log P > -4.2)$ and reduced \chisqr\ of the no-Gaussian fit of $\chisqr_{\rm red_{null}}>2.256\ (\log P < -17)$.
A fit that meets these thresholds is reported, as is the 64 element array,
as unsigned 8 bit values renormalized to a maximum of 255.  

Because the reduced {$\chi^2$} probability of the
Gaussian fit is always near 1 due to the threshold applied,
we define the probability score of a Gaussian $D$ to be \begin{equation}
{S_{\rm Gaussian}(D) = Q(\chi^2_{{\rm red}_{\rm null}}(D)|63) = Q(\frac{63}{2},\frac{\chi_{\rm red_{\rm null}}^2(D)}{2})}
\end{equation}
 
\subsection{\label{sec:folding}Pulses}
A folding algorithm divides a time series into segments of duration equal to the 
period $p$ being searched, and co-adds them in order to improve the signal-to-noise 
ratio for pulses of that period.
The folding algorithm used in SETI@home is a departure from the standard fast folding algorithm (FFA) \citep{staelin69}.
Typical home computer systems at the time the algorithm was developed had small data caches (32 kiB-256 kiB).
A cache miss typically resulted in tens to hundreds of CPU cycles waiting for memory access, whereas a {floating-point} addition would typically complete in one or two cycles.
We found that the standard FFA did not perform well on small-cache machines, and {we} implemented a folding algorithm with the goal of fitting the working set into cache as quickly as possible, at the cost of additional {floating-point} additions.  The benefit of this method, relative to the standard FFA, may not have lasted for more than one generation of microprocessors as larger multilevel caches became the norm.

The client passes the folding algorithm a segment of a PvT array of length ($N$) equivalent to the {half-power} beam crossing time or 40960 time samples, whichever is smaller.  The 40960 sample limit is chosen so {that the} maximum size of the array following the first fold $(N/3)$ will be 64 kiB or less. Subsequent segments are overlapped by 50\% of the array length to ensure maximum sensitivity.  The folding algorithm divides this segment into three equal parts and 
co-adds the data (period $p=\frac{N}{3}$).  The algorithm searches the co-added data for
any peaks above a dynamically computed threshold.
The co-added data {is} 
further divided into two, and again co-added ($p=\frac{N}{6}$) and searched for 
{above-threshold} events.
This process of halving the period is repeated until a period of two
samples is reached.  

The algorithm then returns to the original data {segment} and again divides the
data into three, this time with the upper endpoint of the divided arrays shifted downward by one sample {to}
achieve $p=\frac{N-1}{3}$, and the folding process is repeated.
Once $p=\frac{N}{4}$ is {reached,} the entire segment is again searched, this time folded by four with endpoint shifts until $p=\frac{N}{5}$ is reached.  This repeats for $p=\frac{N}{5}$ to $p=\frac{N}{6}$.  

This results in the following periods being searched.
\begin{align}
p=&{N \over 4 \cdot 2^n} \text{ to } {N \over 3\cdot 2^n},\Delta p={1\over 3 \cdot 2^n}  \\
&\text{ with } n=0 \text{ to }\log_2(\frac{N}{3})-1 \nonumber \\
p=&{N \over 5 \cdot 2^n} \text{ to } {N \over 4\cdot 2^n}, \Delta p={1\over 4 \cdot 2^n} \nonumber \\
  &\text{ with } n=0 \text{ to }\log_2(\frac{N}{4})-1 \nonumber \\
p=&{N \over 6 \cdot 2^n} \text{ to } {N \over 5\cdot 2^n}, \Delta p={1\over 5 \cdot 2^n} \nonumber \\
&\text{ with } n=0 \text{ to } \log_2(\frac{N}{5})-1 \nonumber \\
\end{align}

For the longest segment used (40960 samples),  321\,611 periods between 2 and 13653.3 samples are searched.  {For the shortest DFT used (8) this corresponds to periods between 0.82 ms and 11.2 s.  Periods searched at longer DFT lengths are proportionally longer, with the longest periods becoming limited by either the beam crossing time or the duration of the workunit.}
The periods searched are roughly uniform in logarithmic space,
with a fractional spacing approaching $\frac{1}{N}$.

In principle, further periods missed in this search could be examined.
The primary benefit of this would be increased sensitivity to pulses at these missed periods with pulse duration that is small compared to the duration of a single sample.
However, this would be of limited benefit because the SETI@home client's baseline smoothing removes any signal
with a bandwidth greater than 2 kHz or, equivalently, of duration less than
0.5 ms (\S\ref{sec:algo}).

The probability {that} a time sample $D$ exceed{s} a power threshold $T$ in a noise-like input array of length $N=mn$ {that} has been folded $n$ times to length $m$ is
\begin{equation}
\mathcal{P}(>T)=mQ(n,n\frac{P(D)}{\langle P\rangle })
\end{equation} 
where $\frac{P(D)}{\langle P\rangle }$ is the {pulse power} relative to the mean power in the folded array.  To obtain an equal probability of a false alarm in any element of the folded array regardless of the length of the folded {array,} we chose a constant threshold o{f} $\frac{P}{m}$. 

{Therefore,} we define the pulse probability score
\begin{equation}
S_{\rm pulse}(D)=Q(n,n\frac{P(D)}{\langle P\rangle }).
\end{equation}
Because the length of the searched array (and therefore the number of periods searched) depends upon the rate of motion of the telescope,
a variable power threshold is used to {achieve} a false alarm rate of one detection
per workunit.

The approximate number of power bins searched per workunit is 
\begin{equation}
N_{\rm searched}\left(\dot{\theta}\right)\sim~{6\tten{10}} \frac{\dot{\theta}_{o}}{\dot{\theta}}
\end{equation}
{w}here $\dot{\theta}$ is the rate of motion of the field of view and $\dot{\theta}_{o}$ is the sidereal rate.
In order to obtain a single pulse due to random noise in a standard sidereal rate workunit, we use a motion corrected threshold of
\begin{equation}
{\mathcal T}\left(\frac{P}{m}\right)=\frac{N_{\rm searched}(\dot{\theta})}{N_{\rm searched}(\dot{\theta_{o}})} {\mathcal T}_o
\end{equation}
where ${\mathcal T}_o$ is the threshold at the sidereal rate, computed to result in an average of one detection in a noise-like workunit.

\subsection{\label{sec:triplet}Triplets}

A triplet is defined {as} three events above a threshold evenly spaced in time.
\citet{Dreher_triplet} suggested to us a simple and efficient method {for} finding evenly spaced pulses {in the} data.  Like pulse finding, triplet finding operates on a Doppler-corrected, {single-frequency,} power versus time array.
The array is thresholded at a multiple of the mean noise {power,} and if two or more bins are above a threshold, the bins at the midpoint between each pair of above threshold bins is checked to see if it is also above a threshold.
In {principle,} the midpoint threshold could be different {from} the basic threshold.
{However, in practice,} we found very little difference in overall sensitivity resulting from lowering the midpoint threshold.

The triplet finding algorithm uses as input the same PvT array segments as the pulse finding algorithm sized to match the beam crossing time with subsequent segments overlapping by 50\% to maximize the likelihood that a triplet will be {contained completely} within a segment.  Because the number of segments searched depends
on the telescope motion, we modify the threshold based on the number of unique possible triplets in a workunit, resulting in a threshold $T \propto \frac{\dot{\theta}_{o}}{\dot{\theta}}$.
{This gives a} false alarm probability {of about one per workunit regardless
of the telescope motion.}  The triplet power threshold for workunits acquired when the telescope was moving at
the sidereal rate was approximately $9.0\times$ the mean noise power.

As expected from its construction from three spike-like {signals,} each with two degrees of freedom,
the distribution of triplets in noise-like data is well described by a \chisqr\ distribution with 6 degrees of freedom.
Therefore, we define the probability score of a triplet, $D$, as \begin{equation}
{S_{\rm triplet}(D) = Q(\chisqr|6)=Q(3,\frac{P(D)}{2\langle P\rangle})}
\end{equation}
where $\frac{P(D)}{\langle P\rangle}$ is the average power of the triplet peaks as a multiple of the mean noise power.

\subsection{\label{sec:autocorr}Autocorrelations}

\citet{harp11}~propose that an extraterrestrial
civilization could send a beacon that contains information (and {therefore has an appreciable bandwidth}) but is easily detectable.
This could be done by
sending a signal and then, after a short delay, starting the broadcast
of a copy of the signal.
A signal of this type can be detected by autocorrelation,
which will show a peak power at the given delay.
Once the delay is known, the information within the signal can, in principle,
be recovered.

The client contains an autocorrelation
detector {that examines} delays up to $\pm 64$ki samples ($\pm 6.7$ seconds).
{Following the generation of a power spectrum using a 128ki-point DFT, the client performs a 128\,ki-point inverse transform to compute the autocorrelation.
The autocorrelation function is implemented as 
\begin{align}
F(\nu)&=\mathcal{DFT}_{\rm 128ki}(x(t)) \\
P(\nu)&=\left| F(\nu) F^*(\nu) \right| \nonumber \\
A(\tau)&=\mathcal{DFT}_{\rm 128ki}^{-1}(P(\nu)) \nonumber\\
a(\tau)&=\left| A(\tau) A^*(\tau) \right | \nonumber
\end{align}
where $x(t)$ is the complex time series of the input data.
$P(\nu)$ is the 128\,ki-point power spectrum used to search for spikes
in the 128\,ki-point DFT.}
Because this inverse transform operates on the power spectrum (i.e. the magnitude of the complex {spectrum}), it cannot distinguish between positive and negative delays.

The threshold used for autocorrelation detection is 17.8 times the mean noise power
(following the autocorrelation step). 
As with spikes, Gaussian noise will {result in} a $\chisqr$\  distribution with 2 degrees of freedom, and therefore we use the same
probability score: \begin{equation}
{S_{\rm autocorr}(D)=e^{-\frac{a(D)}{\langle a\rangle }}.}
\end{equation}

\section{\label{sec:analysis}Data analysis}

The SETI@home client takes as input a workunit:
\wuduration of data in a \wubandwidth subband.
It returns a list of detections of the types described above.
We now describe the client algorithm.

\subsection{\label{sec:coherent}Coherent integration}

When searching for narrowband signals, it is best to use a narrow search window (or channel) around a given topocentric frequency.
The wider the channel, the more broad band noise is included in addition to any signal.
This broadband noise limits the sensitivity of the system.
Most recent radio SETI spectrometers have channel widths between 0.5 and 3.0 Hz \citep{SETIBURST, FASTSETI, breakthroughlisten}.

{However, there are limitations to the use of} narrow frequency channels.
One limitation is that extraterrestrial signals are likely to {vary} in topocentric frequency {because of} accelerations of the transmitter and receiver.
For example, a receiver located on the surface of {Earth undergoes an} acceleration
of up to 3.4 \cmpersecsqr due to Earth's rotation. 
{At 1.4 GHz t}his corresponds to a Doppler drift rate of -0.16 \hzpersec.
If {not corrected} for this drift, a transmission
at a constant frequency in an inertial frame
would move outside of a 1 Hz channel in about 6 seconds,
limiting the maximum {coherent} integration time to 6 seconds.
Because of the inverse relationship between maximum frequency resolution
and integration time ($\Delta\nu=\frac{1}{\Delta t}$)
the frequency resolution that can be
effectively used without correcting the received signal for this acceleration is limited to $\Delta\nu\sim$0.4 Hz.

In principle a correction can be made for most of the drift due to motions of the earth, but how does one correct for motions of {a transmitter on or orbiting} an unknown planet?
A transmitter beaming signals directly {at} Earth could correct the outgoing signal for the motions of the transmitter,
but making {such an} adjustment with an omnidirectional beacon is difficult.
Therefore, to search for this type of signal at very narrow bandwidth ($\ll$1 Hz) and with the highest possible sensitivity,
the correction for Doppler drift must be made at the receiving end.
A search for {such} signals must be performed at multiple Doppler drift rates.

It is possible to perform a search for drifting signals using incoherent drift correction.
However, the drift of signals from a frequency channel limits the lossless drift rate to less than $\dot{\nu} \lt \frac{\Delta \nu_{\rm DFT}}{\Delta {t_{\rm int}}}$, where $\Delta \nu_{\rm DFT}$ is the DFT resolution and $\Delta t_{\rm int}$ is the integration time.
This is equal to $\Delta \nu_{\rm DFT}^2$ when a single DFT is considered.  Therefore a typical 1 \Hz spectrometer
begins to lose sensitivity to signals at drift rates greater than 1 \hzpersec.  Spectrometers that sum multiple DFTs into a single spectrum,
such as Breakthrough Listen, 
have the disadvantage of larger $\Delta {t_{\rm int}}$ and a correspondingly smaller lossless drift range. \citep{margot21}

The SETI@home client performs its most sensitive search of the data for signals at drift rates below $\pm$50 \hzpersec (accelerations expected on a rapidly rotating planet) in steps as small as 0.0009 Hz/sec.
This drift rate step is chosen to limit the drift to within a frequency channel over the course of a maximum signal integration time. 

The client examines the data at Doppler drift rates out to $\pm100$ \hzpersec (accelerations of the magnitude that would arise from a satellite in low orbit about a super-earth), but at a more coarse step of 0.015 \hzpersec.
This results in a lower overall sensitivity at these larger drift rates.
In total, as many as \nchirps drift rates are searched in a given workunit.

A signal from a transmitter located on a rotating alien planet would be most likely to {have a} negative drift rate,
as the accelerations involved would be away from the observer.
Positive drift rates could result if a transmitter is in orbit about a planet or star that leaves the transmitter visible while it is being accelerated toward the observer.
{Therefore,} we examine both positive and negative drift rates.
This also leaves open the possibility of detecting a extraterrestrial signal that is transmitted with varying frequency.

{When reporting signals of all types, we report topocentric drift rates.
Corrections of drift rate or frequency to the barycenter are performed in post-processing (see \citeauthor{results_paper} \citeyear{results_paper}).}

\subsection{The analysis algorithm\label{sec:algo}}

\begin{figure}[tb]
\begin{algorithm}[H]
	\begin{algorithmic}
	\State baseline smooth {the data}
	\For{each Doppler rate R}
	    \State frequency drift correct data by R
	    \For{each DFT length L=8 .. 128 ki}
            \For{each {sequential data segment} of length L}
	            \State compute DFT
                \State search DFT for spikes
                \State if (L = 128 ki) look for autocorrelations
                \State store DFT in 2D Power vs Frequency and Time array
            \EndFor    
            \For{each frequency bin in the 2D Array}
	            \State look for Gaussians
	            \State look for triplets
	            \State look for pulses
            \EndFor
	    \EndFor   
	\EndFor
	\end{algorithmic}
\end{algorithm}
\caption{Simplified pseudo code describing the client data analysis.  {Its input is 1 Mi samples of time-domain data.\label{fig:algo}}}

\end{figure}

{The SETI@home client algorithm is summarized in Fig.~\ref{fig:algo}.}
To analyze a workunit, the client first performs a baseline smoothing on the data to remove any wideband ($\Delta\nu > 2$ kHz) features.  {Because the \HI~line is within the recorded band, a value was chosen that will remove even the narrowest \HI~feature, \HI~neutral self-absorption (HINSA), which can have line widths as low as $\sim 0.5\,\kmpersec$ or 2.4\,kHz \citep{goldsmith05}.}
This prevents the client from confusing fluctuations in broadband noise (due in part to variations in the {hydrogen} line emission as the field of view transits the sky) with ETI signals.

{The baseline smoothed is performed on chunks of 32ki complex data points.  A 32ki point power spectrum is computed, and each input point is normalized to the mean power in a 2ki point boxcar around the point.}

The client then loops over a range of Doppler drift rates as described in
the previous section.
At each Doppler drift rate,
power versus time and frequency data cubes are built using DFT of power of 2 lengths, 2$^3 \leq L_{\rm DFT} \leq 2^{17}$ samples;
these result in channel bandwidths of \maxbin to \mbox{\minbin.}
To avoid redundant work, a data cube for a given DFT length is only created when the Doppler drift has changed by an amount that is significant when compared to $1/\Delta\nu^2$.
Therefore, the highest spectral resolution cubes are generated 4 times as frequently as the next higher spectral resolution.
{In the following,} we will refer to a single time row, at all frequencies,
of a data cube as a {\em power spectrum},
and the time series {in} a single frequency 
bin as {\em power versus time} or PvT.

The general Doppler drift correction method creates a reference signal,
$x_{\rm ref}$, with the desired rate of frequency drift,
$\frac{\Delta \nu}{\Delta t}$,
sampled at the same rate as the data.  
\begin{equation}
 x_{\rm ref} = e^{i \phi(t) } = e^{i \pi \frac{\Delta \nu}{\Delta t}t^2} 
\end{equation}
This reference signal is mixed with (i.e. multiplied with the conjugate of) the data, $x$, to derive the drift corrected data.
\begin{equation}
 x_{\rm dedrift} = x_{\rm ref} \overline{x} 
\end{equation}
The signal is corrected incrementally from one drift rate to the next, to limit {the} recalculation of $x_{\rm ref}$.
{We} compute ($\phi(t) \mod 2\pi$) in the exponent incrementally to avoid {errors due to} large sin and cos arguments.

A workunit containing strong RFI can result in {a} large number of detections.  {We determined through observation of early results that the vast majority of workunits containing more than 8 spikes or 30 total detections were contaminated with strong RFI.}
To avoid filling the database with these,
we limit the number of detections returned per workunit
to 8 spikes and 30 total detections.
If either limit is exceeded, the computation is {aborted}
and the detections found up to that point are returned.
A median of 3\% of workunits resulted in this type of overload, although at times of high interference {the fraction of} overloads could reach 30\%. 

\subsection{Sensitivity}

Most radio SETI projects share the same general structure:
a {\em front end} that computes DFTs of time-domain data
and reports {\em events} whose power exceeds a threshold,
and a {\em back-end} that removes RFI and looks for
{\em candidates} for reobservation.

We distinguish two measures of sensitivity for such projects.
{\em Event sensitivity} is the flux level above which an ideal signal
(usually a constant-frequency sine wave) 
will be detected with probability above some threshold,
and {\em detection sensitivity} is the flux level above which
an actual signal (with drift, RFI, and nonzero bandwidth)
will be found as a candidate,
with probability above some threshold.
{Both depend on a number of factors, and either one can be greater.
For the purpose of discovering ET signals, the
detection sensitivity is the relevant measure.
This is discussed in more detail in \citep{results_paper}.}

{In this section, we provide the event sensitivity for the SETI@home analysis of
data obtained at Arecibo.  Sensitivity for observations at other telescopes will
be provided in future publications.}

\subsubsection{Spike Sensitivity}
The received power sensitivity of a single polarization DFT based spectrometer to frequency stable signals much narrower than the channel width is
\begin{equation}
    \Smin({\rm spike})=\mathcal{T}_{\sigma}\frac{2 k_{\rm B} \left(1+\frac{\mathcal{T}_{\sigma}}{l_{\rm DFT}}\right) \left( T_{\rm sys}+T_{\rm sky}\right)}{\Aeff} \sqrt{\frac{\Delta \nu_{\rm DFT}}{ \Delta t_{\rm int}}}\label{eqn:sens}
\end{equation}
where $\mathcal{T}_\sigma$ is the event detection threshold in sigma, $T_{\rm sys}$ is the system temperature in Kelvin ($25-29$K typ., \citeauthor{perillatwebsite} \citeyear{perillatwebsite}), $T_{\rm sky}$ is the sky brightness temperature at the observation frequency and $\Aeff$ is the effective area of the telescope.  The sky continuum brightness temperature is variable from $\sim 3.3$ K near the Galactic poles to $\sim 70K$ near the Galactic plane \citep{galactic_continuum}.  In the narrow frequency range of the hydrogen line, peak \HI\ brightness temperatures of over 150K can be found in the Galactic plane, which would reduce sensitivity in those ranges.  We {used 4K (as noted by \citeauthor{galactic_continuum} \citeyear{galactic_continuum}) as the} median brightness temperature of the Arecibo sky {and a $T_{\rm sys}$ of 29K} to arrive at a median sensitivity calculations.

We derive the effective area of the telescope as
\begin{equation}
\Aeff = \eta_{\rm Q} \eta_{\rm ch} \frac{2 k_{\rm B} \Gamma_{\rm B}}{10^{-26}\,{\rm W\,m^{-2}\,Hz}} \sim  10\,900\,{\rm m^2}
\end{equation} 
where $\eta_{\rm Q}$ is the product of the quantization efficiencies in two-bit complex recording (0.69) and conversion to 4 bit complex data (0.86) \citep{vanvleck},  $\eta_{\rm ch}$ is the mean response of a DFT channel to a signal (0.77 for extremely narrow signals), and $\Gamma_{\rm B}$ is the gain of the outer ALFA receivers in Kelvin per Jansky ($11\,{\rm K\,Jy^{-1}}$ for the inner beam and $8.6\,{\rm K\,Jy^{-1}}$ for the outer beams of the array). Because ${\Delta t_{\rm int}}=\frac{1}{\Delta \nu_{\rm DFT}}$ for spikes, numerically this reduces to
\begin{equation}
\Smin({\rm spike}) \sim \left( 7.9\tten{-26} \joulepermsqr \right) \Delta\nu_{\rm DFT} \mathcal{T}_{\sigma}.
\end{equation}

Because ${\Delta t_{\rm int}}=\frac{1}{\Delta \nu_{\rm DFT}}$ for spikes, the resulting sensitivity in the narrow band limit is $1.4\times{10^{-25}}  \wattpermsqr$ for 128ki DFT lengths ($\Delta \nu_{\rm DFT}=0.075$ Hz).  
For signals of finite bandwidth, both \Aeff and \Smin become functions of the convolution of the signal amplitude with the DFT bin response.  To avoid added {complexity,} we have chosen a Gaussian power {vs.} frequency profile for subsequent analysis.  

\begin{figure}[tb]
\centerline{
\includegraphics[width=\columnwidth]{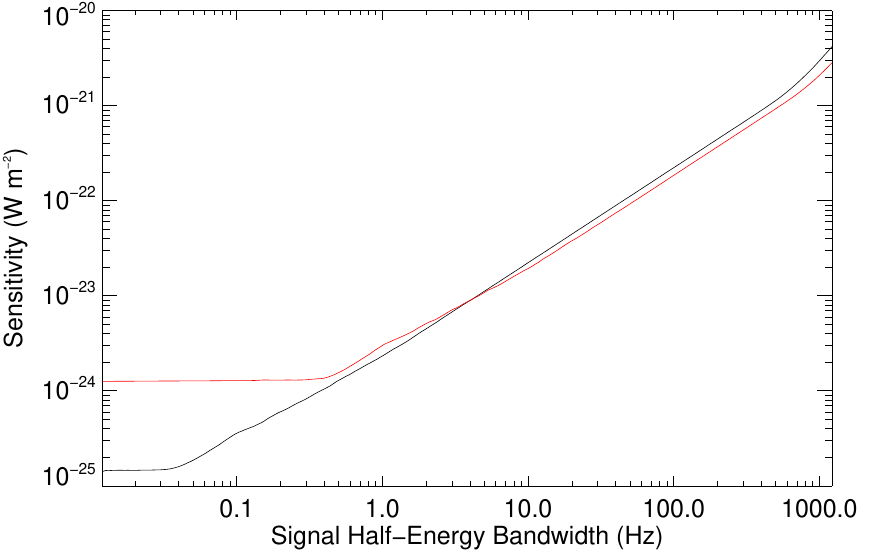}
}
\caption{Comparison of the sensitivity to signals of non-zero bandwidth of SETI@home (black) versus the 0.8 Hz resolution spectrometer SERENDIP VI (red), which also used the ALFA receiver at Arecibo.\label{fig:sens_bw}}
\end{figure}

Fig.~\ref{fig:sens_bw} shows the sensitivity of SETI@home to simulated signals with Gaussian frequency profiles (black) versus the 0.8 Hz resolution spectrometer SERENDIP VI (red), which was also located at Arecibo. For these {signals,} the minimum detectable power is roughly proportional {to the} signal bandwidth.  The deviation from linearity is {mainly} due to an increase in the number of channels in which a signal could be detected as the bandwidth of the signal increases.  At signal bandwidths approaching the workunit bandwidth, the total signal power becomes comparable to the noise power resulting in an increase in \Smin above the linear trend.

\begin{figure}[tb]
\centerline{
\includegraphics[width=\columnwidth]{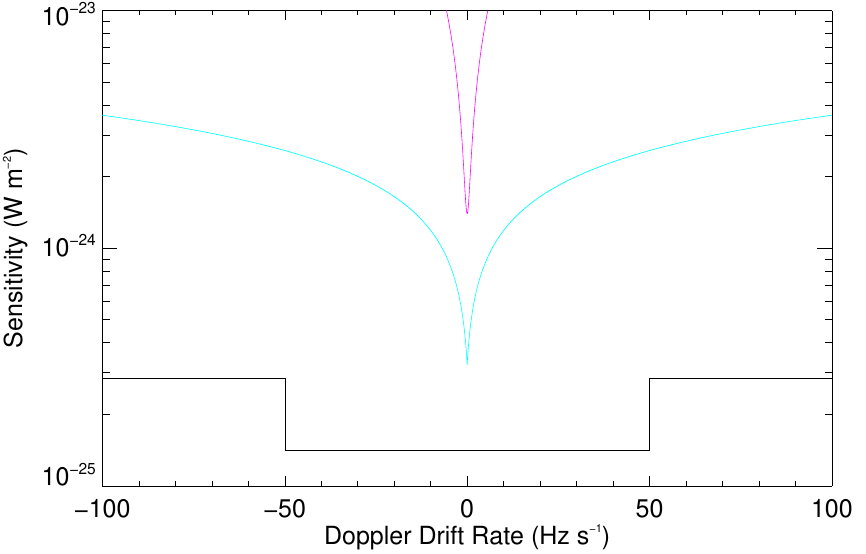}
}
\caption{Comparison of the sensitivity of SETI@home (black) relative to $\sim$0.8 Hz resolution spectrometers using post-threshold Doppler correction (magenta), pre-threshold incoherent Doppler correction (blue) over the frequency drift range $\pm100 \hzpersec$.\label{fig:sens_doppler}}
\end{figure}

{By using} coherent drift correction, SETI@home is able to maintain this sensitivity out to $\pm 50 \hzpersec$ and a sensitivity $4\times$ higher out to $\pm100 \hzpersec$.
Fig.~\ref{fig:sens_doppler} shows the event sensitivity of the SERENDIP VI without pre-threshold Doppler drift correction (magenta) and 30 second integration, a theoretical instrument comparable to SERENDIP VI with pre-threshold incoherent Doppler drift correction (blue), and the approach used in the SETI@home client (black).  

\subsubsection{Gaussian Sensitivity}

Because the Gaussian threshold is set on reduced \chisqr, rather than directly on power, Eqn. \ref{eqn:sens} applies only approximately to the case of the Gaussian detection type. 
\begin{equation}
\Smin({\rm Gaussian}) \sim 1.9\tten{-25} {\Delta \nu_{\rm DFT}\sqrt{\frac{\dot{\theta}}{\dot{\theta_o}}}} 
\,\wattpermsqr
\end{equation}
Because the Gaussian fit requires 64 bins in the PvT array, a DFT length of 16ki or less is required.  A 16k DFT ($\Delta \nu_{\rm DFT}$=0.6 Hz) results in a sensitivity of 1.1\tten{-25} \wattpermsqr when the beam is traversing at the sidereal rate.

\subsubsection{Triplet Sensitivity}
With pulsed signal {types,} there are multiple ways of expressing the sensitivity.  We could express triplet sensitivity as the received power while the signal is on, similar to spikes
\begin{equation}
\Smin({\rm triplet})=\mathcal{T}_{\sigma}\frac{2 k_{\rm B} \left(1+\frac{\mathcal{T}_{\sigma}}{l_{\rm DFT}}\right) \left( T_{\rm sys}+T_{\rm sky}\right)}{\Aeff} \Delta \nu_{\rm DFT}\label{eqn:tripsens}
\end{equation}
with $\mathcal{T}_{\sigma}\sim 9$ for the sidereal beam transit rate, resulting in a sensitivity of $\Smin \sim 5.25 \tten{-26} \wattpermsqr$ for the finest frequency resolutions.  However, this might be misleading as there is a far larger parameter space available for triplets at coarser resolutions.

We could express the sensitivity as the total energy flux received in a single pulse.  In that case,
\begin{equation}
\Emin({\rm triplet})=\Smin({\rm triplet}) \Delta t_{\rm DFT} \sim \left( 7.9\tten{-26} \joulepermsqr \right) \mathcal{T}_{\sigma} 
\end{equation}
or $7.1\tten{-25} \joulepermsqr$ at the sidereal transit rate.  This method has the advantage of being independent of the resolution at which the pulse is detected.

Finally, we can express the threshold at the average power over the {entire pulse period} $p$, which is indicative of the energy requirements for transmission.  In this case,
\begin{equation}
\langle\Smin({\rm triplet})\rangle=\frac{\Emin({\rm triplet})}{p}
\end{equation}
We prefer this notation because, for a given power budget a low duty cycle pulse can potentially be detected at a greater distance than one of high duty cycle of equivalent averaged power. There is a limit to the benefit of short pulses {because, as the} pulse bandwidth increases, interstellar dispersion becomes increasingly important.  As described in \S\ref{sec:dispersion} pulse durations shorter than about 50 \us, interstellar dispersion becomes important at Galactic distances, which could limit the effectiveness of $\Delta\nu\Delta t=1$ pulses as an interstellar beacon.

\subsubsection{Pulse Sensitivity{\label{sec:pulse_sensitivity}}}

\begin{figure}[tbp]
\centerline{
\includegraphics[width=\columnwidth]{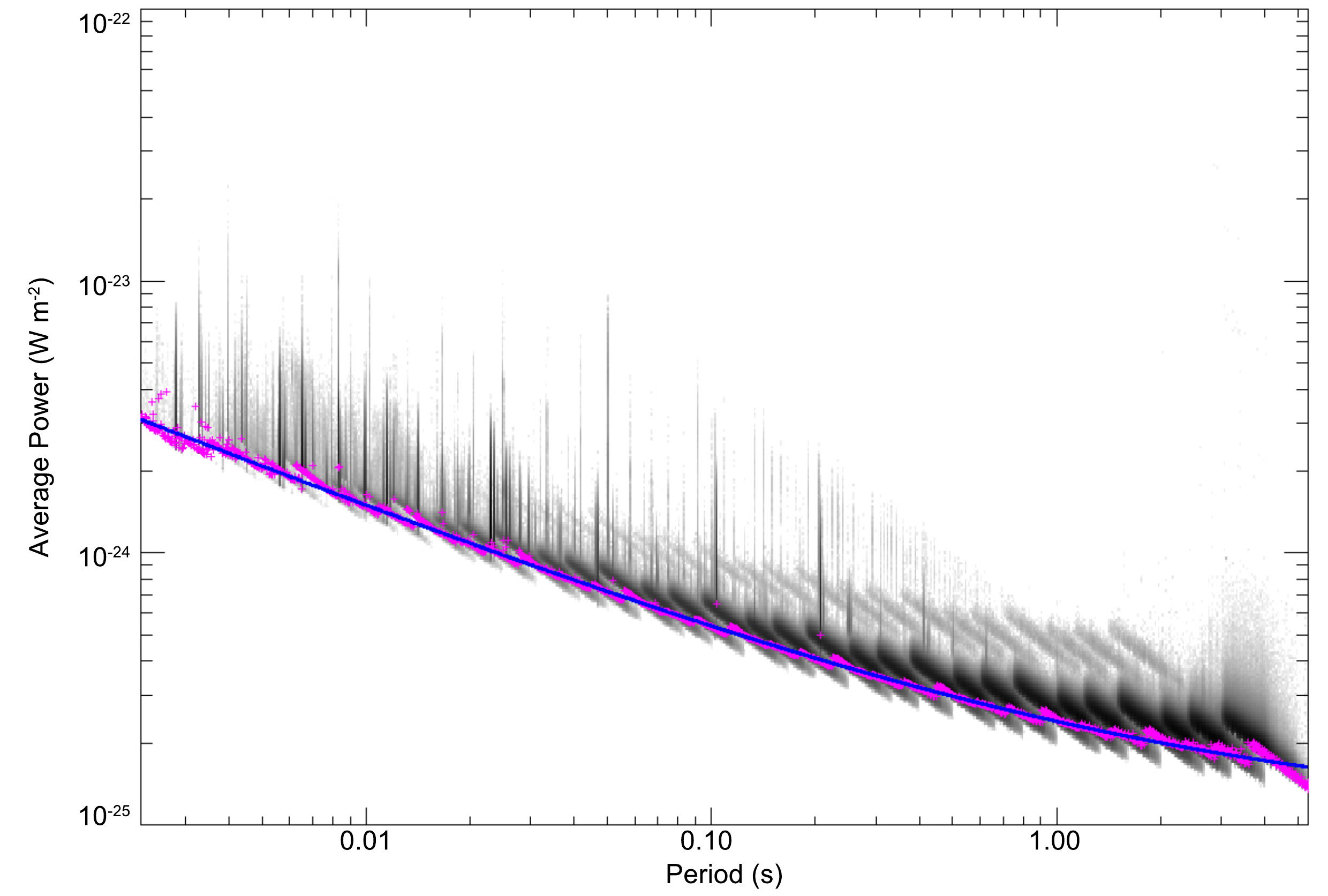}
}
\caption{Average power versus pulse period for {140 million} pulses detected by SETI@home {at times the telescope beam was moving within 5\% of the sidereal rate. The magenta points mark the 5th percentile, which provides an estimate of sensitivity to pulses of that period.  The blue line is a fit to those points,
as described in \S\ref{sec:pulse_sensitivity}.} 
\label{fig:pulse_sens} }
\end{figure}

The constant false alarm probability threshold for pulses, combined with their large parameter space, makes it difficult to express the sensitivity in terms of a function of beam crossing time, period, and search bandwidth without calculating the inverse of the incomplete Gamma function.  {Instead,} we look at the average power distribution of pulses detected in {noise-like} data. Fig.~\ref{fig:pulse_sens} shows the average power and period of {1.4\tten{8}} pulses detected by SETI@home {at times when the telescope beam was moving within 5\% of the sidereal rate.   Noiselike detections are found in a band}, extending from 3\tten{-24} \wattpermsqr at a {period of 2.2 ms} to {1.6}\tten{-25}\,\wattpermsqr at a period of 5.34 seconds. {The vetical lines present in the image show common RFI features and represent about 18\% of the signals. 
 The diagonal features are noiselike detections that roughly follow loci of constant pulse amplitude.  Because every range of periods is examined at multiple bandwidths and multiple numbers of folds, it is difficult to provide a single number expressing the sensitivity at a given period.
 To estimate the sensitivity at any small range of period we calculated the 5th percentile of pulse power.  We then fit these points with a smooth function to provide a heuristic estimate of sensitivity.  For periods between 2.2 ms and 5.34 seconds our detection sensitivity is}
\begin{equation}
\langle\Smin({\rm pulse})\rangle=1.4\tten{-24}\,\wattpermsqr \left(\frac{0.01\, \s}{p}\right)^{0.51} + 1.1\tten{-25}\,\wattpermsqr.
\end{equation}

\subsubsection{Autocorrelation Sensitivity}
Sensitivity to autocorrelation signals reverts to the standard form of Eqn.~\ref{eqn:sens}, with an additional $\sqrt{2}$ due to the folding of positive and negative correlations into a single bin.  {The u}se of a lower threshold {achieves} nearly {the same} sensitivity {as for} spikes.  Because autocorrelation is {performed only} at the finest DFT resolution (128 ki), a single value can express this sensitivity. 
\[
\Smin({\rm autocorr})=\mathcal{T}_{\sigma}\frac{2 k_{\rm B} \left( T_{\rm sys}+T_{\rm sky}\right)}{\Aeff} \Delta \nu_{\rm DFT}=1.5\tten{-25}\, \wattpermsqr
\]
{
\section{\label{sec:testing}Testing and validation}

The SETI@home front end consists of three main parts:

\begin{itemize}
\item The data recorder takes analog signals from 14 ALFA feeds
and metadata from the observatory (time and alt/az pointing).
It outputs files containing digital data in a 2.5 MHz band
and sampled metadata.
\item The splitter takes these files
and outputs workunits, each comprising 107s of data in a 9.7 KHz band
and sampled metadata with pointings in RA/dec.
\item The client takes these workunits and outputs detections,
whose attributes include time, freq, power, Doppler drift rate, and sky position.
\end{itemize}

We conducted several tests to validate these parts,
individually and in combination.

First, we validated the client's algorithms for finding all detection types
by generating synthetic workunit files,
each containing a target signal embedded in Gaussian noise.
The signals consisted of a chirped sine wave, pulsed with a given period and duty cycle,
with a Gaussian envelope corresponding to a given telescope motion.
The workunits contained pointing and timing data consistent with this motion.
We generated these workunits with signals sampling the full
range of target signal parameters and with various simulated telescope slew rates.
We processed each workunit with the client
and verified that the output included detections whose parameters
matched those of the synthetic signals.

Second, we validated the splitter by generating synthetic full-band data
with signals embedded in noise,
splitting them, processing the resulting workunits with the client,
and checking its output.

Third, we validated the full system
(including telescope electronics, data recorder, splitter,
and the client's spike detection) by injecting an RF sinusoid into the telescope.
An oscillator phased locked to the observatory's
hydrogen maser frequency reference is used to generate a stable
sinusoid of known frequency, time, and power.
This signal is transmitted via a small antenna that viewed the ALFA receiver through
a hole in the Arecibo primary reflector.
We verifed that the result of processing the corresponding workunits
includes spikes at the appropriate frequency, time, and power.
This provides a basic test of the feed, the low-noise amplifer,
receiver electronics, analog mixers, filters, and amplifiers,
the SETI@home data recorder, the SETI@home splitter,
and the SETI@home front-end analysis program.

\begin{figure*}[tbp]
\centerline{
\includegraphics[width=2\columnwidth]{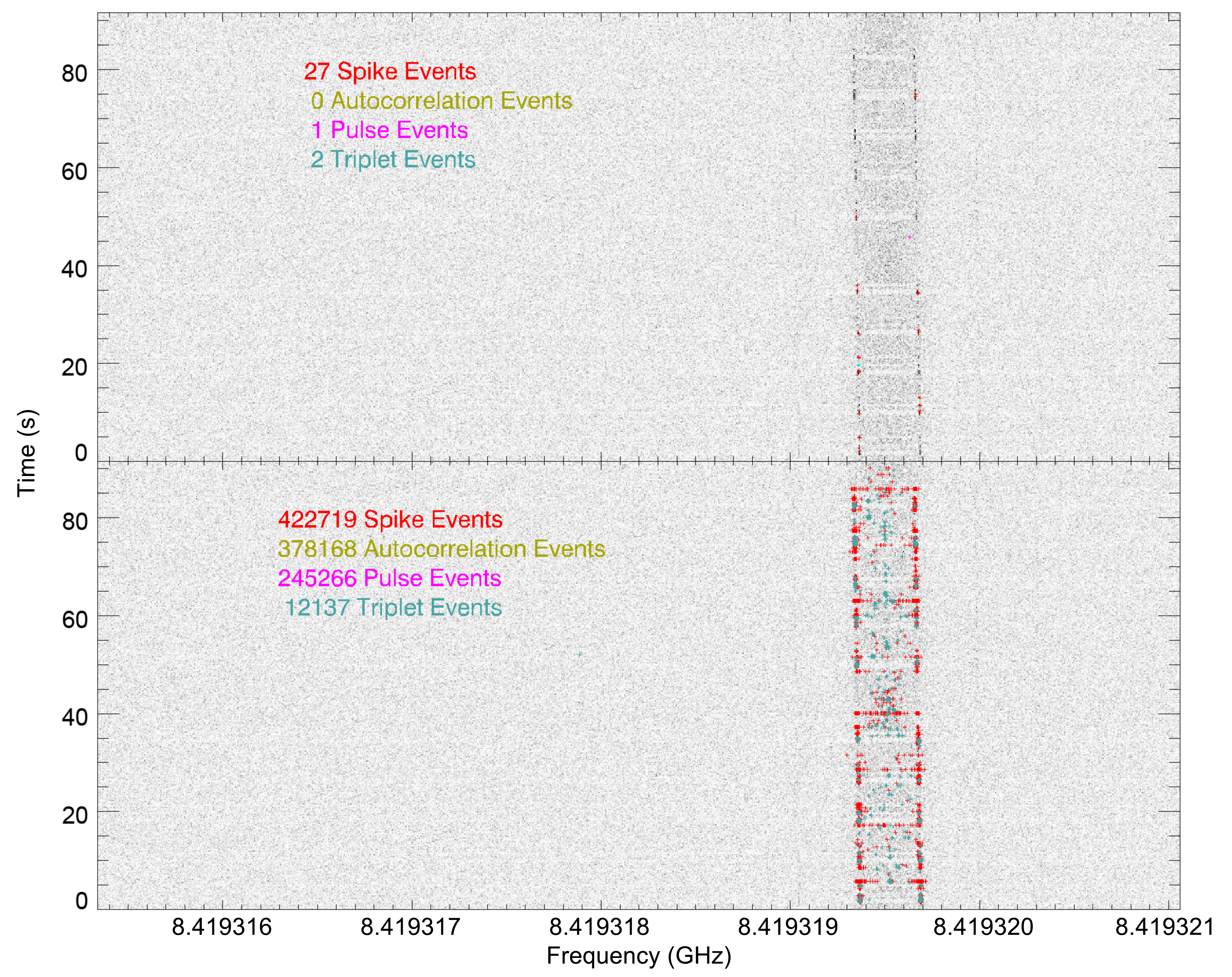}
}
\caption{\label{fig:voyager}Waterfall plots showing the SETI@home analysis of a GBT observation of Voyager 1.  The upper panel shows the results with the default cutoff of 30 total detections.  The lower panel show detections when the analysis is allowed to run to completion.}
\end{figure*}

Fourth, we validated the splitter and client by successfully detecting
an on-sky signal from the Voyager 1 spacecraft.
This data was provided to us by the Breakthrough Listen project.
It was recorded at the Green Bank telescope
at 1916 UTC on 2016 Sept 19 (JD 2457651.324) while tracking the spacecraft.
The client generated spike, pulse, and autocorrelation detections with the expected
frequency, position, and Doppler drift rate.
This verified that the splitter and client were handling frequency correctly.

These results are shown in Fig.~\ref{fig:voyager}.
The upper and lower frames show the same grayscale image
of power versus frequency and time
generated using sequential DFTs of the workunit data.
The frequency modulated data channel is visible on the right side.
Detections are displayed using color coding,
with spikes in red and triplets in cyan.
Because the telescope is tracking,
no Gaussian search is performed.
Pulses and autocorrelation are not shown in this plot.
The full duration of the workunit is folded, hence all pulses have the same time.
Because autocorrelations use the entire bandwidth of the workunit,
their frequency is the midpoint of the band corrected for Doppler drift.
The upper panel shows the result with the default SETI@home parameters,
aborting after 30 detections are found.
The lower panel shows the result if the analysis is allowed to run to completion:
1\,058\,290 detections throughout the time range in four detection types.  

Despite being the most distant anthropogenic radio source,
the Voyager 1 data channel is quite strong,
with detections at more than 100$\times$ the mean noise power.
In the complete analysis, the peak of the spike power distribution versus Doppler drift
rate is at a Doppler drift rate of -0.370 \hzpersec
which closely corresponds to the barycentric Doppler drift rate of the observation.
The power distribution of spikes found in the Voyager carrier
also peaked at the correct barycentric drift rate with powers
of up to 2800$\times$ the mean noise power.

Finally, we validated the handling of time and pointing information
by the data recorder, splitter and client.
To do this, we recorded data from tracking and drifting observations of the Crab pulsar.
We split the data and analyzed the workunits
with both the SETI@home client and the Astropulse client,
which uses the same position and time determination code as SETI@home.
The Astropulse client detected pulses with the correct sky position.
All of these were 'giant' pulses; we verified their times by
comparing the Astropulse output with that of a wideband spectrometer
operated by the observatory.
The SETI@home client, as expected,
did not detect the pulses because of its removal of wideband features.
However, because the average power in the continuum increased greatly
when the telescope was pointed at the pulsar,
the signal-to-noise ratio of the RFI features present in the SETI@home data decreased.
We were able to use this effect to validate the client's time and position determination,
as well as the $\sigma$-width used for our Gaussian fitting.
}

\section{\label{sec:volunteer}Using volunteer computing} 

{SETI@home uses {\em volunteer computing} to perform front-end data analysis.
Volunteers install a {\em client program} on their computing devices
(home computers and smartphones).
The program fetches jobs from a central server and processes them.
It has an optional screensaver function that shows a visualization
of the analysis.
}

{We} initially developed {our} own client/server software for volunteer
computing functions: job distribution, screensaver logic, {etc.}
This system required volunteers to install a new version of
the program each time our data analysis algorithm changed.

In 2005 {we} moved to the Berkeley Open Infrastructure for Network Computing (BOINC) platform \citep{Anderson20}
for volunteer computing,
which allows algorithm updates without user involvement.
BOINC has been used for projects in many science areas,
such as climate research, drug discovery,
cosmology, pulsar and gravitational wave detection, and number theory.
Volunteers can install the BOINC client program on their computers
and configure it to contribute computing power
to any or all of these projects.

The use of volunteer computing {provided} a large amount of computing power.
However, it introduced a number of {issues, which are} described below.

\subsection{Volunteer recruitment and retention}

SETI@home was launched in May 1999,
and for about a year it received considerable worldwide media coverage.
This produced a surge of volunteers,
peaking at about one million active participants.
{After the} media coverage subsided,
the volunteer population gradually declined.

We developed, in collaboration with the BOINC project,
a number of mechanisms designed to attract new volunteers and retain existing ones.
Some of these mechanisms required administration..
When {possible,} we used existing volunteers for these purposes:

\begin{itemize}

\item Technical support for new volunteers was provided
{using} online message boards; experienced volunteers
answered questions posed by new volunteers.
We also developed a system where one-on-one technical
support was provided via Skype.
\item We operated message boards for volunteers
to discuss science, computing, and other topics.
We used volunteers to moderate these message boards,
suppress{ing} spam and ``flame wars".
\item We provided web-based ``leader boards" {listing}
the volunteers who provided the most computing power.
This motivated some volunteers to run SETI@home on more
computers, and in some cases to buy powerful, multi-GPU computers
for the purpose of running SETI@home.
\item We created a system {that allows} volunteers to form teams,
typically based on nationality, institution, or computer type.
{We added l}eader boards {for} teams.
This motivated some volunteers to recruit friends and family
to {boost} their team statistics.

\end{itemize}

Studies have shown that participants in volunteer computing
and other forms of ``citizen science" have several motivations
\citep{strasser23, nov14}.
These include competition and community,
as well as support for science goals.
The mechanisms listed above were designed to
support these various motivations.

\subsection{Device heterogeneity}

The pool of volunteered computers was varied \citep{Anderson09, korpela12a}.
{The computers had various processor types (Intel, ARM),
bitness (32- and 64-bit),
CPU features (such as SSE3 and AVX2), and number of cores.
They had different} operating systems
(Windows, Mac OS, Linux, Android)
and different versions of these.
Many computers had Graphics Processing Units (GPUs)
{capable of general-purpose computing.}
These GPUs had different makers (NVIDIA, AMD, Intel),
different models, and different driver versions.

{Getting as much computing power as possible from a given computer
typically required one or more versions of the client program:
for example, a GPU version for that particular GPU model
and a CPU version to use the remaining CPU time.
We developed dozens of such versions,
trying to fully exploit as wide a range of computers as possible.
BOINC provides features that automatically select
the best-performing versions for a given computer.}

Volunteers {assisted in these efforts.}
BOINC has a feature called ``anonymous platform"
that allows volunteers to use their own
{client} versions.
Volunteers used this to develop versions optimized
for particular CPU {features}
and to develop versions that use GPUs.
In many {cases,} we eventually added these to the set of official versions.
Volunteers also restructured {many} algorithms in the SETI@home client
to improve {their speed and} numerical accuracy.

\subsection{Result verification}

Results returned by volunteer computers may be incorrect
for a variety of reasons:
hardware errors due to overclocking and overheating,
bugs in particular application versions,
and in some cases hacking by volunteers trying to get
credit for jobs not actually performed.

BOINC provides a mechanism for detecting incorrect results using replication.
Each job is executed on two different computers,
and the results are accepted only if they agree;
{otherwise,} the job is run on a third computer,
and so on until a consensus is reached.
This mechanism worked well.
However, different processor types and numerical libraries
typically differ in the low-order bits of floating-point calculations,
and these deviations accumulate in calculations such as DFT.
Thus, in comparing the results of two replicas of a job,
we tolerate a certain amount of variation.
This depends on the parameter: for example,
frequencies {must} agree within .1 Hz,
while parameters like power must have a relative difference
of at most 1\%.

In its original form, replication resulted in a 50\% loss in effective computing power.
To reduce this overhead, the mechanism was refined so that
computers {that} return {several} consecutive verified results
are gradually exempted from replication.  {These trusted
computers would still be randomly sent some replicated results as a check.  If result verification failed either because of a mismatch, or because of values outside of the range of valid calculations, the computer would be marked as untrusted until it had returned a number of valid results}.
This reduced the overhead to a few percent.

\subsection{Server and network performance}

We had to implement various server functions:
web server, job scheduler, data splitter,
file download and upload, database servers, and so on.
At first, we divided these functions {between} three desktop computers.
These were quickly overwhelmed and we moved to a collection of
dedicated server computers and network storage devices,
eventually numbering 20 or so.

Initially, these servers were located {at} our research center,
whose Internet connection {provided} 100 Mbps in each direction.
Our network traffic - primarily sending workunits - saturated that,
and we had to rent a commercial 1 Gbps connection.
Later, we moved our server complex to the UC Berkeley campus hosting facility,
which provided ample network capacity.

\subsection{Computing power}

\begin{figure}[tbp]
\centerline{
\includegraphics[width=\columnwidth]{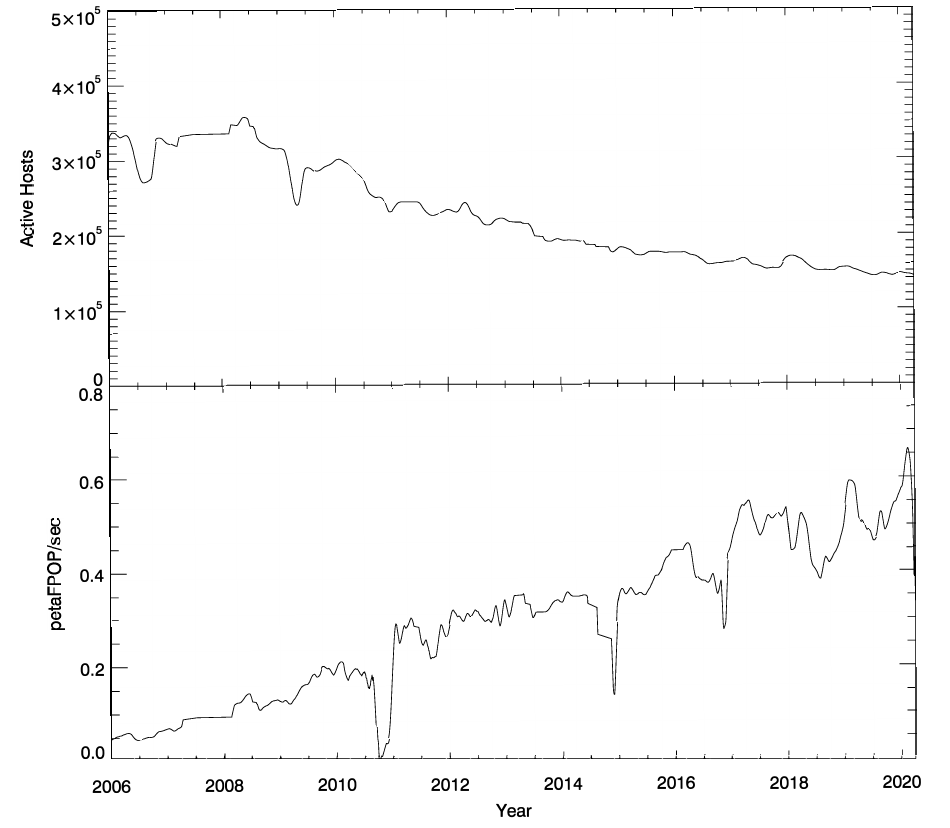}
}
\caption{The upper panel shows the number of computers actively participating in SETI@home.  The lower panel shows the rate of computing done by these machines in petaFPOP/s ($1$ 
petaFPOP=$10^{15}$ floating point operations).}
\label{fig:comp_power}
\end{figure}

The computing throughput of SETI@home varied over time,
as shown in {Fig.~\ref{fig:comp_power}.}
This variation is due to several factors.
Between 2006 and 2020 the number of {computers actively participating decreased} from 350\,000 to 140\,000.
However, the average floating-point performance of
the computers grew at a greater rate,
due to increases in CPU clock rate and number of processors,
and (starting in about 2010) the introduction of graphics processing units
(GPUs) capable of general-purpose floating-point computing
at speeds one or two orders of magnitude faster than CPUs.
{Thus,} the computing throughput grew from about 100 teraFPOP/s in 2006
to 600 teraFPOP/s in 2020.
In total, SETI@home used roughly
$6 \cdot 10^{23}$ floating-point operations.

{In a typical workunit, about half the computing
(in terms of floating-point operations)
went to computing DFTs,
and about half to the fast folding algorithm for pulses.
The time for other functions, such as Gaussian and triplet finding,
was small in comparison.

Table \ref{table:platforms} breaks down computing
power by operating system.
Table \ref{table:windows} compares the major client
versions for Windows.
CUDA is a library for NVIDIA GPUs,
while OpenCL is a cross-platform GPU library.
For GPU versions, CPU time is typically less than elapsed time
because the CPU often has to wait for the GPU.
On a given computer, BOINC chooses versions
based on available hardware and software.
On computers with GPUs, it typically runs both CPU and GPU versions
in order to fully utilize the processing resources.

\begin{table}[tbp]
\begin{center}
\begin{tabular}{l l}
Operating system & Fraction of computing \\
\hline
Microsoft Windows & 71.2\% \\
Apple MacOS & 15.8\% \\
Linux & 12.6\% \\
Android & 0.4\% \\

\end{tabular}
\caption{SETI@home computing power by operating system.\label{table:platforms}}
\end{center}
\end{table}

\begin{table*}[tbp]
\begin{center}
\begin{tabular}{l l l l}
Version & Fraction of computing & Median CPU time (s) & Median elapsed time (s)\\
\hline
NVIDIA GPU, OpenCL & 70.9\% & 519 & 548 \\
CPU & 13.6\% & 7746 & 8229 \\
AMD GPU, OpenCL & 7.9\% & 182 & 626 \\
NVIDIA GPU, CUDA & 7.6\% & 206 & 2012 \\
\end{tabular}
\caption{Comparison of SETI@home client versions for Windows.}
\label{table:windows}
\end{center}
\end{table*}
}

\section{\label{sec:related} Related Work and Contributions of SETI@home}

Several radio SETI project have surveyed large sky areas.
Some were {\em commensal}, collecting data while
telescope pointing was being controlled by other projects.
These include searches at the Hat Creek and Green Bank observatories
\citep{werthimer88}
and at Arecibo
\citep{cobb00},
\citep{bowyer16}.
Other projects have done sky surveys using dedicated telescopes.
These include the early Ohio State
project and its ``Wow!'' signal
\citep{kraus77}
as well as the ``Fly's Eye" project
at the Allen Telescope Array
\citep{siemion12}.

In addition, there have been a number of targeted searches
that observed particular stars (and sometimes galaxies).
These include OZMA and OZMA II at Green Bank
\citep{drake60, sagan75, drake86, gray21},
Phoenix at Arecibo and ATA \citep{backus02},
and Breakthrough Listen projects at Parkes and Green Bank
\citep{price20, enriquez17}.
Observations of 33 stars were {recently made} at
the FAST observatory in China \citep{luan23}.

The projects differ in sky coverage,
frequency coverage, and sensitivity.
Table \ref{tab:surveys} shows parameters of some of the projects.
\begin{table*}[tbp]
\centering
\caption{Parameters of three sky surveys and three targeted searches (adapted from \citet{Wright18}).  Event sensitivity is relative to constant-frequency sinusoids.\label{tab:surveys}}
\begin{tabular}{ccccccc}
\hline
Telescope & Project & Sky& Event  & Frequency & Bandwidths & Signal Drift\\
 &  & Coverage & Sensitivity & Coverage & Searched & Rate Coverage\\
 & & degree$^2$ & $10^{-26} \wattpermsqr$ &  MHz   &  Hz  & $\hzpersec$ \\
\hline
\hline
Arecibo	&  SETI@home &  12\,000	& 14 &  2.5 & 0.07 - 1220 & $\pm$100
\\
Arecibo & SERENDIP VI\footnote{\citet{SETIBURST}} & 12\,000	&  110  &  280	&  0.8 &  $\pm$0.6  
\\
MWA	& \footnote{\citet{Tingay16}}	&   400	 & 50	&     24	&  10\,000 &  $\pm$1
\\
ATA &	ExoplanetNHZ\footnote{\citet{Harp16}} &   8	&   265	 &  2000  &    0.7	 & $\pm$1
\\
Arecibo &	Phoenix\footnote{\citet{backus02}}	&  0.3	&  16 &   1250	&  1   &  $\pm$1
\\
Arecibo	 & Listen\footnote{\citet{Enriquez18}} &    11 &	46	 & 800	& 2.7, 1000 & $\pm$7
\\\hline
\end{tabular}

\end{table*}

This table shows that, compared to other sky surveys,
SETI@home has better event sensitivity but smaller
frequency coverage.
However, SETI@home differs from previous radio SETI projects in ways
that are not shown in the table.\vskip 0.5em

{\noindent \bf Multiple time and frequency resolutions:}
SETI@home analyzed data at 15 octaves of time and frequency resolution,
ranging from \minbin (\maxtimebin) to \maxbin (\mintimebin).
Other SETI projects have used only one or two different spectral and time resolutions  \citep{harp18,breakthroughlisten}.
The use of multiple resolutions improves sensitivity to both narrowband
signals and pulsed signals.\vskip 0.5em

{\noindent \bf Coherent integration at a wide range of Doppler drift rates:}
SETI@home was the first project to use coherent integration,
increasing its sensitivity to narrowband signals.
SETI@home used coherent integration
at \nchirps drift rates from \minchirp\ to \mbox{\maxchirp,}
to compensate for transmitter acceleration
at a range of possible planetary or orbital parameters.
Recently, other projects have used coherent integration,
but only to compensate for receiver acceleration
due to the Earth's motion
\citep{margot23, horowitz93}.\vskip 0.5em

\noindent{\bf Multiple detection types:}
Observations in which a Gaussian beam moves across a point source
would be expected to produce a Gaussian-shaped power curve.
Of projects with moving beams,
SETI@home was the first to search for such patterns.
It was also the first SETI project to search
for pulsed signals using a folding algorithm,
to search for triplets \citep{Dreher_triplet},
and to search for autocorrelations.
The idea of searching for autocorrelations
was proposed by \citet{harp11} and later implemented
in a search at the Allen Telescope Array \citep{harp18}.\vskip 0.5em

These differences involve the SETI@home front end.
In addition, the SETI@home back-end has a number
of features that are unique among existing projects:
for example, its use of {\em candidate birdies}
(which are used to evaluate RFI algorithms
and to estimate candidate sensitivity)
and its ability to find signals whose Doppler shift
changes by large amounts (on the order of 100KHz)
over long time periods; see
\cite{results_paper}.

\section{\label{sec:conclusion}Conclusion and future work}

We have described the goals and architecture of SETI@home
and have presented the details of its front end,
which uses volunteer computing to analyze time-domain data
and identify five types of detections.
Volunteer computing allowed us to use coherent integration
for increased sensitivity to narrowband signals,
and to detect pulsed signals, Gaussians, and autocorrelations.

The back-end of SETI@home takes this set of detections,
removes RFI, and identifies sets of detections that are consistent
with having a single persistent source.
We used the back-end to a) identify a set of 200 signal candidates,
which we are reobserving at the FAST observatory \citep{FAST2019},
and b) estimate the {\em candidate sensitivity} or the sensitivity of the system as a whole to detecting candidates.
The back end and its results are described in a companion paper \citep{results_paper}.

There are several ways in which the front end of SETI@home
(or similar future projects) might be improved:

\begin{itemize}

\item Include multiple (or all) beams in each workunit.
If a detection is just above threshold in one beam,
it may match with detections just below threshold in others.
This would {allow us to} lower the thresholds and find weaker signals.
It could also help {identify} RFI before it enters the detection database.
Because we limit the number of signals returned from each workunit,
this would also result in increased sensitivity to weaker signals.

\item Include the signals from both polarizations of each beam in a workunit.
This would increase sensitivity
and would allow a Stokes parameter search for circularly polarized emissions,
which are a theorized means of creating an identifiable and detectable technosignature \citep{circular_pol}.

\item Eliminate the spike detection type, using only Gaussians to detect {continuous} narrowband signals.
When the telescope is moving, a celestial source would produce a near-Gaussian power envelope {because of} the shape of the telescope beam;
when the telescope is not change{moving,} the envelope would be constant,
which one can think of as an infinite-width Gaussian.
Any series of spikes not matching such an envelope is unlikely to be from a celestial source.
It would also be possible to enforce a Gaussian envelope for pulsed and autocorrelation signals.

\item {Complete the analysis at both zero Doppler drift rate and at the barycentic drift rate, even in the presence of strong RFI.
Aborting the analysis before reaching the barycentric drift rate could cause very strong extraterrestrial signals to be misidentified as RFI by the back end.}

\item Do Gaussian fitting based on momentary telescope motion rather than the average motion over the workunit.
This would provide greater sensitivity when the telescope motion is changing during the workunit.

\end{itemize}

SETI@home has demonstrated the viability of using volunteer computing for radio SETI front-end processing.
Future projects could use this approach
in combination with radio telescopes such as FAST \citep{FAST2019}
or array telescopes such as the Allen Telescope Array \citep{ATA2009}
{or the} Square Kilometer Array \citep{dewdney09}.

SETI@home was designed in the early 2000s,
when Pentium chips and dial-up Internet connections were common.
This constrained parameters such as frequency coverage.
Since then, consumer technology has evolved in many dimensions,
leading to new possibilities for future ratio SETI projects
using volunteer computing.

The processing power of home computers -- especially their GPUs -- continues to increase.
There are several billion such computers,
and with appropriate promotion and {incentives,}
it may be
possible to harness many millions of them.
So future projects may have far more computing power than SETI@home.
This will make it feasible to analyze larger frequency {ranges}
and may enable new detection methods such as the
Karhunen-Lo\`{e}ve transform \citep{KLT}.

The speed of home Internet connections has increased
to the 1 Gbps range,
and the free disk space on a typical home computer
has grown to the Terabyte range.
These trends will reduce potential bottlenecks in
{the analysis of} larger frequency ranges.

\subsection{Acknowledgements}

Millions of volunteers contributed to SETI@home by donating
the processing power of their {computers} and cell phones.
Many volunteers helped in other ways, as described in \S\ref{sec:volunteer}.
{We especially thank Raistmer, Josef~W.~Segur, Jason Groothuis, Urs Echternacht, David Woolley, Tetsuji Rai, Charlie Fenton, Richard Haselgrove, Byron Leigh Hatch, Takuya Ooura,  Matteo Frigo, Steven G. Johnson and Roelof Engelbrecht for their assistance with algorithms, debugging, porting, optimization, and other useful discussions.
Willy de Zutter supplied data related to computing power}.

SETI@home has been supported by grants from Starwave, The Planetary Society,
the state of California, National Science Foundation grant 1407804, 
the Marilyn and Watson Alberts SETI Chair fund,
and by donations from individuals.
We received equipment donations from Sun Microsystems, Intel, NVIDIA, AMD/Xilinx, NetApp,
Hewlett Packard, Fujitsu, Quantum, Seagate, Western Digital,
and Packet Clearing House.
We used data storage resources {at} the National Energy Research
Scientific Computing Center, a DOE Office of Science User Facility
supported by Contract No. DE-AC02-05CH11231.
SETI@home's observations were made at the NAIC Arecibo Observatory,
a facility funded by the NSF and operated by Cornell University {and the University of Central Florida.}

\bibliographystyle{aasjournal}
\bibliography{sah}
\end{document}